\documentclass[twocolumn,aps,amsmath,amssymb,showpacs,nofootinbib,preprintnumbers,prd]{revtex4-1}
\usepackage[utf8x]{inputenc}
\usepackage{graphicx}
\usepackage{enumitem}
\usepackage{slashed}
\usepackage[T1]{fontenc}
\usepackage{url}
\usepackage[pdftex]{hyperref}
\usepackage[T1]{fontenc}
\usepackage{bm}
\usepackage{xcolor}
\usepackage{color}
\usepackage{amsmath}
\usepackage[export]{adjustbox}
\usepackage[normalem]{ulem}
\usepackage[caption=false]{subfig}
\usepackage{float}
\usepackage{hyperref}
\usepackage{soul}
\usepackage[capitalise]{cleveref}
\crefformat{equation}{Equation~(#2#1#3)} 
\crefformat{figure}{Figure~(#2#1#3)}
\crefformat{section}{Section~(#2#1#3)}
\crefformat{table}{Table~(#2#1#3)}
\crefformat{appendix}{Appendix~(#2#1#3)}
\usepackage{comment}

\hypersetup{
    colorlinks=true,
    linkcolor=red,
    filecolor=cyan,      
    urlcolor=blue,
    citecolor=blue
    }

\usepackage{diagbox}

\usepackage{tikz}
\usepackage[compat=1.1.0]{tikz-feynman}
\usetikzlibrary{quantikz}
\tikzset{every node/.append style={font=\small}}
\tikzset{>=latex}
\tikzset{every picture/.append style={line cap=round, line join=round}}
\tikzset{every path/.append style={>=latex, line width=0.8pt}}

\renewcommand\[{\begin{equation}}
\renewcommand\]{\end{equation}}

\newcommand{\LL}[2]{L_{#1#2}}
\newcommand{\QL}[2]{Q_{#1#2}}
\newcommand{\uR}[1]{u_{R#1}}

\newcommand{\chis}[1]{\chi_{_{#1}}}
\newcommand{\ychi}[1]{y_{_{#1}}}
\newcommand{\mylength}{1.3cm}

\DeclareUnicodeCharacter{2212}{\ensuremath{-}}
\begin{document}

\preprint{PITT-PACC-2507}
\preprint{MI-HET-865}

\title{A Baryon and Lepton Number Violation Model Testable at the LHC}

\author{Amit Bhoonah}

\email{amit.bhoonah@pitt.edu}

\affiliation{Pittsburgh Particle Physics, Astrophysics, and Cosmology Center, Department of Physics and Astronomy, University of Pittsburgh, Pittsburgh, USA}
\author{Francis Burk}

\email{frb25@pitt.edu@pitt.edu}

\affiliation{Pittsburgh Particle Physics, Astrophysics, and Cosmology Center, Department of Physics and Astronomy, University of Pittsburgh, Pittsburgh, USA}

\author{Da Liu}

\email{liudaphysics@gmail.com}

\affiliation{Pittsburgh Particle Physics, Astrophysics, and Cosmology Center, Department of Physics and Astronomy, University of Pittsburgh, Pittsburgh, USA}

\author{Tong Ou}

\email{tongou@uchicago.edu}

\affiliation{Enrico Fermi Institute, Department of Physics, University of Chicago, Chicago, Illinois, 60637, USA}

\author{Deepak Sathyan}

\email{dsathyan@tamu.edu}

\affiliation{Maryland Center for Fundamental Physics, Department of Physics, University of Maryland, College Park, MD 20742, USA}

\affiliation{Mitchell Institute for Fundamental Physics and Astronomy, Department of Physics and Astronomy, Texas A\&M University, College Station, TX 77843, USA}

\begin{abstract}

Proton decay experiments typically constrain baryon number violation to the scale of grand unified theories. From a phenomenological point of view, this makes direct probing of the associated new resonances, such as the X and Y bosons, out of reach for even the most optimistic future experiments. It has, however, been known that certain specific patterns of baryon and lepton number violation can suppress proton decay by multiple powers of the masses of the heavy resonances involved, opening the possibility that the observed limits on the proton lifetime 
are consistent with baryon number violating physics at energy scales much lower than that of grand unification. We construct an explicit example of such a model which violates baryon number by one unit, $\Delta \text{B} = -1$, and lepton number by three units, $\Delta \text{L} = -3$, and show that despite stringent limits on the predicted $p \rightarrow e^{+}/\mu^{+} \overline{\nu}\overline{\nu}$ mode from the Super-Kamiokande experiment, the masses of the newly introduced elementary particles can be $\mathcal{O}$(TeV). We identify interesting unique signatures of baryon number violation of this model that can be probed both with currently available LHC data and with the upcoming High-Luminosity LHC. We also present a scenario for low-scale baryogenesis within the framework of this model. 
\end{abstract}

\maketitle

\section{\label{sec:Introduction}Introduction}
The conservation of baryon (B) and lepton (L) number in the observed interactions of elementary particles is an established experimental fact. From a theoretical point of view, however, this raises many questions as there are no fundamental principles at play behind the existence of either of these global symmetries. Moreover, there are many reasons to consider them to be accidental effects, valid only up to the roughly 10 TeV energy scale that has been directly accessed so far in experiments. For example, it is predicted that global symmetries are explicitly broken by quantum gravitational effects \cite{Abbott:1989jw,Kallosh:1995hi,Lee:1988ge}. 
The validity of this argument in the absence of an experimentally verified quantum theory of gravity is an open question. Yet, based on robust theoretical foundations and experimental observations, there are compelling arguments in favour of a more complete theory of nature with additional elementary particles whose interactions violate either \emph{or both} B and L.
Namely, baryon number violation is one of three necessary conditions for realizing \textit{baryogenesis} \cite{Sakharov:1967dj}, and lepton number violation is common in mechanisms that generate neutrino masses, which are necessary to explain the observed flavour oscillations in the neutrino sector \cite{Davis:1968cp, Super-Kamiokande:1998kpq, SNO:2001kpb}. \\

The most distinctive prediction of most theories with Baryon Number Violation (BNV) is, of course, that the proton --- or a bound neutron --- should decay. This is the subject of a rich experimental program, with modes such as $p \rightarrow \pi_{0} \ e^{+}/\mu^{+}$ and $p \rightarrow K^{+} \overline{\nu}$ having been extensively searched for and, so far, not observed \cite{Super-Kamiokande:2020wjk,McGrew:1999nd,Kamiokande-II:1989avz,Super-Kamiokande:2014otb,Super-Kamiokande:2005lev}. A variety of more exotic nucleon decay modes have also been considered --- see \cite{Heeck:2019kgr} for a summary --- but, even in these channels, experiments have yet to identify conclusive evidence for BNV. These null results put stringent constraints on the masses of new particles that violate B and L, potentially pushing them to values above $10^{15}$ GeV, a characteristic scale of Grand Unified Theories (GUTs) \cite{Georgi:1974sy,Pati:1974yy,Fritzsch:1974nn}. \\

In light of this, one might be tempted to associate any BNV interaction with very high energy scales, inaccessible to even the most optimistic of future collider experiments. There is, however, one caveat to these conclusions: GUT theories typically violate B and L by one unit each, keeping the difference, B -- L, as a conserved quantity. This in turn induces low energy BNV operators at dimension six. Given a measured proton lifetime of $\tau_{P}$ years, one can argue by naive dimensional analysis that the scale associated with BNV is roughly
\begin{equation*}
 M_{\text{BNV}} \sim \left(5000\times \frac{\tau_{P}}{10^{32} \ \text{yrs}} \right)^{\frac{1}{4}}  \times 10^{15} \ \text{GeV},  
\end{equation*}
where we take the proton mass to be 1 GeV for simplicity and neglect phase space factors. \\

It has, however, been shown that if instead of B -- L, the underlying theory of nature\footnote{Perhaps, more accurately, an intermediate version of it valid for energies above the electroweak scale but below that of Grand Unification.} conserves a more general linear combination B + aL, a  $\in$ $\mathbb{Q}$, the BNV effective operator is generated at a dimension higher than six \cite{Weinberg:1980bf,Wilczek:1979hc}. This causes a suppression of proton-decay-inducing operators by multiple powers of $M_{\text{BNV}}$.  
In fact, for such a low energy effective operator of dimension $d>6$ and a corresponding proton lifetime $\tau_{P}$, 
\begin{equation*}\label{eq:ndagenerald}
    M_{\text{BNV}} \sim \left(5000\times \frac{\tau_{P}}{10^{32} \ \text{yrs}} \right)^{\frac{1}{4}}  \times \left(10^{15}\right)^{\frac{4}{2d-8}} \ \text{GeV},  
\end{equation*}
once again neglecting phase space factors and taking the proton mass to be 1 GeV. From this estimate, for sufficiently high $d$, the mass scale of new particles involved in BNV interactions can lie significantly below the GUT scale. As an example, $d$ = 11 sets $M_{\text{BNV}} \approx 10$ TeV, within reach of current or near-term future collider experiments such as the Large Hadron Collider (LHC) and its upcoming High Luminosity (HL) upgrade. \\

For such theories, where $M_{\text{BNV}}$ is low enough to lie within the reach of the LHC or other near-term future colliders, the new resonances responsible for the BNV interactions can be physically produced, and the ultraviolet (UV) complete model can be directly probed.
As we will demonstrate in this work with a specific model, the collider probe can be even more powerful than the experiments that probe proton decay directly, if the proton decay rate predicted from the model is beyond the reach of even upcoming experiments like Hyper-Kamiokande (Hyper-K) \cite{Hyper-Kamiokande:2018ofw}.
Complementary to the collider search, the new particles in the UV complete model can also be probed by low energy precision experiments, as they will generally have B and L conserving couplings to the known elementary particles of nature. 

Such experiments include searching for flavour violation in the quark and leptonic sectors, measuring rates of rare meson decays, and measuring the anomalous magnetic moments of leptons.
The results to date from these experiments have yielded very stringent constraints on the parameter space of the UV model, which must be taken into account when analyzing the collider search for the new particles. \\
  
In this work, we will illustrate the points made above using a theory that violates B and L by one and three units, respectively, but conserves B -- L/3,\footnote{The conserved B -- L/3 global symmetry is anomalous, but since we make no attempt to gauge it, this poses no serious implications for the model itself.} or 
\begin{equation*}
\Delta \text{B}  = \frac{1}{3}\Delta \text{L}= -1.     
\end{equation*}
In this scenario the proton is predicted to decay into three anti-leptons, 
\begin{equation*}
p \rightarrow e^{+}/\mu^{+}\overline{\nu}\overline{\nu},   
\end{equation*}
a process for which the Super-Kamiokande (Super-K) experiment has searched. No evidence of this decay was found, setting a lower limit on the proton lifetime of $\tau_{_{P}} >$ 1.7 $\times 10^{32}$ years if the charged anti-lepton is a positron and $\tau_{_{P}} >$ 2.2 $\times 10^{32}$ years if it is an anti-muon \cite{Super-Kamiokande:2014pqx}. We find that the proton decay rate for this model implies a bound of roughly $M_{\text{BNV}} > $ 1 TeV, opening up the possibility of studying this model at the LHC. Our study shares many similarities with the analysis of \cite{Fonseca:2018ehk}. \\

Interestingly, the model discussed in this work can provide a natural explanation of the baryon asymmetry of the universe, in a mechanism similar to the post-sphaleron baryogenesis developed in \cite{Babu:2006xc, Babu:2013yca}. 
A novel feature of this mechanism is that it utilizes the new high-dimensional BNV interaction for the baryon number generation, instead of the electroweak sphaleron that is known to be the only BNV source in the Standard Model and is commonly used in baryogenesis mechanisms.
To avoid the generated baryon asymmetry being washed out by electroweak sphalerons, we choose an appropriate mass scale for the participating particles such that the baryogenesis is only effective when the electroweak sphalerons have become inactive.
This can be easily achieved thanks to the suppression from the high dimensionality of the BNV interaction.
The baryogenesis process also has to violate CP symmetry according to the Sakharov's conditions \cite{Sakharov:1967dj}, which can be induced purely from the CP violations in the quark and leptonic sectors of the Standard Model, and hence can be directly constrained by future neutrino precision measurements. \\

The remainder of this work is organized as follows: we present the model in question in Section~\eqref{sec:Model}. In Section~\eqref{sec:ExistingPheno} we perform a precise calculation of the proton decay rate which establishes a lower bound of $\mathcal{O}$(TeV) on $M_{\text{BNV}}$, and survey all relevant B and L conserving constraints applicable to our model, after which we propose a dedicated search for its distinctive BNV signature that can be performed at the LHC and HL-LHC in Section~\eqref{sec:collider_pheno}. We present a detailed account of the low scale baryogenesis scenario we just outlined in Section~\eqref{sec:baryogenesis} and conclude in Section~\eqref{sec:conclusion}.

\section{Model $\&$ Effective Operators}\label{sec:Model}
\begin{table}[t!]
    \centering
 \begin{tabular}
 {|c |c |c |c |c |c |}
    \hline
    \textbf{Particle} & $\mathbf{SU(3)}$ & $\mathbf{SU(2)_{L}}$ & $\mathbf{U(1)_{Y}}$ & $\mathbf{U(1)_{B}}$ & $\mathbf{U(1)_{L}}$ \\ 
    \hline
    $\chi_{_{uu}}$ & $\mathbf{3}$ & $\mathbf{1}$ & $-\frac{4}{3}$ & $-\frac{2}{3}$ & 0 \\ 
    $\chi_{_{LQ}}$ & $\mathbf{3}$ & $\mathbf{1}$ &$-\frac{1}{3}$ & $\frac{1}{3}$ & $+1$ \\
    $\chi_{_{LL}}$ & $\mathbf{1}$ & $\mathbf{1}$ & $+1$ & 0 & $-2$ \\

    \hline
    \end{tabular}
    \caption{Quantum Numbers of the three new scalars of the B -- L/3 conserving model considered in this work.}
    \label{tab:ParticleContent13}
\end{table} 
We add three new scalar fields to the Standard Model of Particle Physics (SM). The first is a diquark, $\chi_{_{uu}}$, that couples two quarks and arises in the context of Grand Unification models based on the $\text{E}_{6}$ \cite{Hewett:1988xc} or Pati-Salam \cite{Pati:1974yy,Mohapatra:2007af} gauge groups. In our theory, it couples two right handed up-type quarks of different generations,
\begin{equation}\label{eq:LSDiquark1}
\begin{split}
     \mathcal{L}_{uu} = y^{ij}_{_{uu}}\epsilon^{c_{1}c_{2}c_{3}} \ \overline{u}^{c}_{ic_{1}}P_{R}u_{jc_{3}}\chis{uu c_{2}} + \text{h.c}.
\end{split}
\end{equation}
In the equation above $i$ and $j$ label quark flavours, $i \neq j$  due to fermionic statistics, and the $c$ indices label QCD color. The second particle we add is a leptoquark, $\chi_{_{LQ}}$, also common in many GUT theories --- particularly the ones based on $SU(5)$ \cite{Buchmuller:1986iq} or $SO(10)$ \cite{Senjanovic:1982ex}. These theories predict many different types of leptoquarks, but we will only employ one and couple it to a left-handed quark and a left-handed lepton,
\begin{equation}\label{eq:LLeptoquark1}
\begin{split}
    \mathcal{L}_{_{LQ}} &=y^{ij*}_{_{LQ}} \epsilon^{\alpha\beta}\ \overline{Q}^{c}_{\alpha ic_{1}}\chi^{\dagger c_{1}}_{LQ} L_{\beta j} + \ \text{h.c.}\\
   & =y^{ij*}_{_{LQ}} \ \overline{u}^{c}_{ic_{1}}P_{L}l_{j}\chi^{\dagger c_{1}}_{LQ} - y^{ij*}_{_{LQ}} \ \overline{d}^{c}_{ic_{1}}P_{L}\nu_{j}\chi^{\dagger c_{1}}_{LQ} + \ \text{h.c},
    \end{split}
\end{equation} 
where $Q,L$ are the SM left-handed $SU(2)_L$ quark and lepton doublets correspondingly, and $\alpha,\beta$ are the weak isospin indices.
Finally, we add a Zee-Babu \cite{Zee:1980ai,Babu:1988ki} singly charged dilepton singlet, $\chi_{_{LL}}$, coupling two left handed leptons of different generations (once again due to fermion statistics),
\begin{equation}\label{eq:LSDilepton1}
\begin{split}
    \mathcal{L}_{_{LL}} &= y^{ij}_{_{LL}} \ \epsilon^{\alpha\beta} \overline{L}^{c}_{\alpha i}\chi_{_{LL}}L_{\beta j}  + \  \text{h.c}.\\
    &= y^{ij}_{_{LL}} \ \overline{\nu}^{c}_{i}P_{L}l_{j}\chi_{_{LL}} - y^{ij}_{_{LL}} \ \overline{l}^{c}_{i}P_{L}\nu_{j}\chi_{_{LL}} + \  \text{h.c}.
\end{split}
\end{equation}
This naturally brings up the question of neutrino masses, which can be achieved in exactly the same manner as the Zee-Babu mechanism, i.e., by introducing either a doubly charged scalar or a second Higgs doublet to our model --- we leave this for a dedicated future study. In general, the Yukawa couplings in Equations\,(\ref{eq:LSDiquark1})---(\ref{eq:LSDilepton1}) can be complex numbers, which may lead to interesting CP violation phenomena. We will focus on real Yukawa couplings in the mass basis for this work. \\

Since these Yukawa interactions represent a \emph{hard} breaking of B and L, we must assign quantum numbers to the new scalars such that both global symmetries are conserved, as listed in Table~\ref{tab:ParticleContent13}.

So far, nothing about this model is different from other works that have studied these scalars individually \cite{Pascual-Dias:2020hxo,Giudice:2011ak,Dorsner:2016wpm,Crivellin:2021ejk,Zee:1980ai,Babu:1988ki}, aside from assigning them the correct quantum numbers to enforce B and L conservation. Therefore, these \emph{cannot} induce BNV interactions, even at loop level. To achieve BNV, we introduce the \emph{only} baryon and lepton number violating interaction of this model, one that occurs in the scalar sector:
\begin{equation}\label{eq:BViolatingInteraction}
\mathcal{L}_{\text{BNV}} =  \Lambda_{\text{BNV}} \ \chi^{\dagger}_{_{LQ}}\chi_{_{uu}}\chi_{_{LL}} + \ \text{h.c}.  
\end{equation}
This trilinear interaction breaks B and L softly but conserves B -- L/3. It is worthwhile to take a brief detour from our presentation of this model and speculate a little on the origins of this term. A motivated possibility, from the point of view of baryogenesis that we will discuss later, is that it arises from the interaction of the three $\chi$s with a complex SM singlet scalar field $\Phi$ that acquires a vacuum expectation value (VEV),   
\begin{equation*}
    \lambda\left<\Phi\right> \chi^{\dagger}_{_{LQ}}\chi_{_{uu}}\chi_{_{LL}} \rightarrow \Lambda_{\text{BNV}}\,\chi^{\dagger}_{_{LQ}}\chi_{_{uu}}\chi_{_{LL}}, 
\end{equation*}
where $\lambda$ is a dimensionless coupling constant.
Since $\Phi$ is a complex scalar field, it is always possible to rotate away the B and L violation in this interaction if the latter is not fixed by any other interaction. The theory then violates B and L spontaneously after $\Phi$ acquires a vacuum expectation value, while still preserving B -- L/3.  This, of course, implies  the existence of a massless Goldstone boson that could, for example, add an extra source of radiation which over-closes the universe. To remedy this situation, we can make the global symmetry associated with $\Phi$ only approximate, for example by adding a cubic interaction term of the form $(\Phi^{*}\Phi)\Phi$ to the Lagrangian. Then, one can use the phase rotation to fix either the cubic self-interactions of $\Phi$ or its interactions with the $\chi$s, but not \emph{both}. For example, using the phase rotation to make the quartic interaction with the $\chi$s baryon number conserving effectively assigns to $\Phi$ a baryon and a lepton number that is then softly broken by its cubic self-interactions. Spontaneous symmetry breaking then results in the appearance of a pseudo-Goldstone boson carrying both B and L with interactions with the SM that are potentially weak enough to make it a dark matter candidate. We leave a dedicated study of these interesting cosmological scenarios to future work. \\

The other interactions in the scalar sector can be written compactly as:
\begin{widetext}
\begin{equation}\label{eq:BViolatingInteraction}
\begin{split}
    \mathcal{L}_{S} =  \frac{\mu^{2}_{j}}{2}\chi^{\dagger}_{j}\chi_{j} + \lambda_{jj} \ (\chi^{\dagger}_{j}\chi_{j})^{2} + \lambda_{jk} \ \chi^{\dagger}_{j}\chi_{j}\chi^{\dagger}_{k}\chi_{k}+ \lambda_{jH} \ \chi^{\dagger}_{j}\chi_{j}H^{\dagger}H,
\end{split}
\end{equation}    
\end{widetext}
where $j=LQ,LL,uu$ and, without loss of generality, we assume that $\mu^{2}_{j}$ > 0. This imposes $\lambda_{jj} > 0$ and $\lambda_{jk} > 0$ since the new scalars all carry non-zero electromagnetic charge and therefore cannot acquire vacuum expectation values. Mass positivity imposes a bound on their mixing with the Higgs boson 
\begin{equation}
  \lambda_{jH} > - \frac{\mu^{2}_{j}}{v^{2}} 
\end{equation}
where $v = 246$ GeV is the Higgs VEV. Imposing perturbative unitarity places important theoretical constraints on the parameters of this potential --- particularly for $\Lambda_{\text{BNV}}$ --- as described in \cref{sec:PerturbativeUnitarity}. \\

We close this presentation of our model by mentioning the low energy effective operators that can be derived from these interactions and are important phenomenologically. Of special interest is the one responsible for BNV, which appears at dimension nine,
\begin{equation}\label{eq:BNVdim9}
    \mathcal{O}^{\text{BNV}}_{\text{9}} = \frac{c_{9}}{M^{5}_{\text{BNV}}}\overline{Q}^{c}_{i\beta} L_{j\alpha}\overline{u}^{c}_{k}P_{R}u_{l} \overline{L}^{c}_{m\gamma}L_{n\delta}\epsilon^{\alpha\beta}\epsilon^{\gamma\delta},
\end{equation}
where 
\begin{equation}
\begin{split}
    M^{5}_{\text{BNV}} = \frac{\Lambda_{\text{BNV}}}{\left(m_{\chis{LQ}}m_{\chis{uu}}m_{\chis{LL}}\right)^{2}} \ \text{and} \ c_{9}  = y^{ij}_{LQ}y^{kl}_{uu}y^{mn}_{LL}
\end{split}
\end{equation}
is the Wilson coefficient, $i,j,k,l,m,n$ and $\alpha,\beta,\gamma,\delta$ are the flavour and $SU(2)$ indices respectively, and we have suppressed colour labels for clarity. Since the two right handed up-type quarks resulting from integrating out the diquark have to be of different generations, at least one of them has to be either a charm or a top quark, which kinematically forbids proton decay at tree level but allows it at one-loop through flavour violation in charged current interactions, as we will see in the next section. 

\section{Existing Constraints}\label{sec:ExistingPheno}
Having introduced the model, we now consider its phenomenology. Since our focus is on BNV, the starting point is naturally proton decay: it sets the scale at which new resonances should appear and guides the choice of experiment to test them, if any are available at all. 
In our case, this exercise will show that the scale in question is $\mathcal{O}$(TeV), which opens up a variety of low energy probes of this model that we then survey before discussing their implications on the predicted proton lifetime in our model. 

\subsection{Proton Decay}
\begin{figure}[h!]
\begin{tikzpicture}
    \begin{feynman}
        \vertex [label=left:$u_R(p_3)$](uR);
        \vertex [below = 1.cm of uR, label=left:$d_{L}(p_1)$] (dL);
        \vertex [below = 1.cm of dL, label=left:$u_L(p_2)$] (uL);
        \vertex [right = 1.2cm of uR] (uu1);
        \vertex [right =1.2cm of dL] (w1);
        \vertex [right=1.2cm of uL] (w2);
        \vertex [right = 1.2cm of uu1] (uu2);
        \vertex [right = 1.2cm of uu2] (LL);
        \vertex [above right = 0.5cm and 1.0cm of LL, label=right:$\bar{\nu}_{Lm}(k_2)$] (l2);
        \vertex [below right=0.5cm and 1.0cm of LL, label=right:$l^+_{Ln}(k_3)$] (l3);
        \vertex [right = 1.2cm of w2] (LQ2);
        \vertex [right = 2.2cm of LQ2] (l1) {$\bar{\nu}_{Ll}(k_1)$};

        \diagram*{(uR) --[fermion] (uu1),
        (dL) --[fermion](w1) --[fermion, edge label=$u_k$](uu1),
        (uL) --[fermion](w2) --[fermion, edge label=$d_{Lj}$](LQ2),
        (w1) --[boson, edge label' = $W^{-}$](w2),
        (uu1) --[scalar, edge label=$\chi_{uu}$](uu2) --[scalar, edge label=$\chi_{LL}$](LL),
        (l2) --[fermion](LL),
        (l3) --[fermion] (LL),
        (uu2) --[scalar, edge label=$\chi_{LQ}^\dagger$](LQ2),
        (l1) --[fermion] (LQ2),
        };
    \end{feynman}
    \node[cross out, draw, minimum size=1.5pt, line width=0.6pt] at ($(w1)!0.8!(uu1)$) {};
\end{tikzpicture}
\begin{tikzpicture}
    \begin{feynman}
        \vertex [label=left:$u_R(p_3)$](uR);
        \vertex [below = 1.cm of uR, label=left:$d_{L}(p_1)$] (dL);
        \vertex [below = 1.cm of dL, label=left:$u_L(p_2)$] (uL);
        \vertex [right = 1.2cm of uR] (uu1);
        \vertex [right =1.2cm of dL] (w1);
        \vertex [right=1.2cm of uL] (LQ2);
        \vertex [right = 1.2cm of uu1] (uu2);
        \vertex [right = 1.2cm of uu2] (LL);
        \vertex [above right = 0.5cm and 1.0cm of LL, label=right:$\bar{\nu}_{Lm}(k_2)$] (l2);
        \vertex [below right=0.5cm and 1.0cm of LL, label=right:$l^+_{Ln}(k_3)$] (l3);
        \vertex [right = 1.2cm of LQ2] (w2);
        \vertex [right = 2.2cm of w2] (l1) {$\bar{\nu}_{Ll}(k_1)$};

        \diagram*{(uR) --[fermion] (uu1),
        (dL) --[fermion](w1) --[fermion, edge label=$u_k$](uu1),
        (uL) --[fermion](LQ2),
        (w2) --[fermion, edge label=$l_{Lj}$](LQ2),
        (w1) --[boson, edge label = $W^{-}$, near end](w2),
        (uu1) --[scalar, edge label=$\chi_{uu}$](uu2) --[scalar, edge label=$\chi_{LL}$](LL),
        (l2) --[fermion](LL),
        (l3) --[fermion] (LL),
        (LQ2) --[scalar, edge label'=$\chi_{LQ}^\dagger$, near end](uu2),
        (l1) --[fermion] (w2),
        };
    \end{feynman}
    \node[cross out, draw, minimum size=1.5pt, line width=0.6pt] at ($(w1)!0.8!(uu1)$) {};
\end{tikzpicture}
\caption{Feynman diagrams for loop level proton decay, where $j,k,l,m,n$ are flavour indices. A chirality flip along the $u_k$ propagator is necessary to ensure the correct chirality on both vertices, which is induced by a mass insertion denoted by the $\times$. When the two antineutrinos are of the same flavour, $l = m$, each of these diagrams has a companion with $k_{1} \leftrightarrow k_{2}$. On the upper panel, the $W$ boson couples an incoming up quark and a virtual down type quark of flavour $j$. On the lower panel it couples an outgoing antineutrino and a virtual lepton of flavour $j$. Because of the $SU(2)$ index contraction of the leptoquark couplings, these two diagrams carry a relative minus sign between them and interfere destructively.}
\label{fig:ProtonDecayUVComplete}
\end{figure}
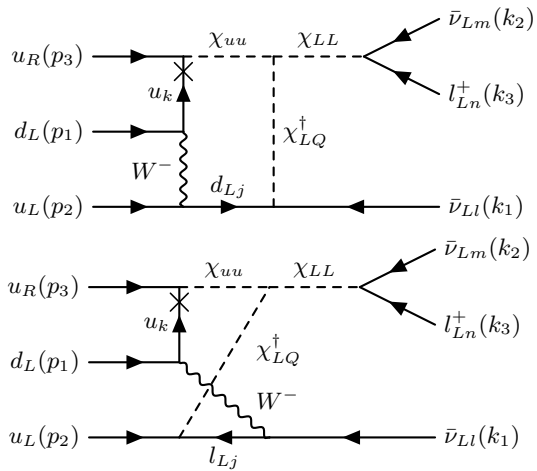

Let us begin our precise evaluation of the proton decay rate, focusing on important conceptual ideas here and leaving a detailed write-up in Appendix~\eqref{sec:PDDetailedCalculations}. There are two relevant Feynman diagrams for the $p\rightarrow \overline{\nu}\overline{\nu}e^{+}/\mu^{+}$ process, as shown in Figure~\eqref{fig:ProtonDecayUVComplete}. If the outgoing antineutrino flavours $l=m$, each of the diagrams will have a companion with the antineutrino momenta interchanged, $k_{1} \leftrightarrow k_{2}$. To proceed, we split the amplitude as 
\begin{equation}
    \mathcal{M}_{p \rightarrow \overline{l}\overline{\nu}\overline{\nu}} = \mathcal{M}^{q}_{p \rightarrow \overline{l}\overline{\nu}\overline{\nu}} + \mathcal{M}^{l}_{p \rightarrow \overline{l}\overline{\nu}\overline{\nu}}, 
\end{equation}
corresponding to the amplitudes for the upper and lower panels of Figure~\eqref{fig:ProtonDecayUVComplete} respectively (explicit expressions for these are given in equations~\eqref{eq:ampprotondecayquark} and~\eqref{eq:ampprotondecaylepton}). The two diagrams carry a relative minus sign between them due to the antisymmetric contraction of $SU(2)$ indices in leptoquark-fermions interactions, causing destructive interference.  \\

The amplitudes are proportional to two quark masses corresponding to the two chirality flips expected from an effective operator analysis \cite{Weinberg:1980bf}. The first of these is from an off-shell charm or top quark interacting with an up quark from the proton. Since this involves a diquark vertex, these are all right handed. However, the $c$ or $t$ must also attach to a left-handed $d$ quark from the proton, so a chirality flip is required, as indicated by the cross in Figure~\eqref{fig:ProtonDecayUVComplete}. The second chirality flip occurs due to the vector nature of the loop integral, which introduces factors of momentum ``slashed'' that can then be applied to quark fields --- this is discussed in detail in Appendix~\eqref{sec:PDDetailedCalculations}.  Importantly, for the second chirality flip, the only non-zero contribution introduces a factor of the up quark mass, but we do not take it to be that of a free up quark as, in this case, it is the valence quark of a proton. Instead, we use the constituent quark mass \cite{Lavelle:1995ty,ParticleDataGroupConstituentQuarks:2024cfk}, which we denote by $m^{*}_{u}$, and set to one third the proton mass,\footnote{The precise value, the determination of which requires detailed lattice calculations, has no appreciable effect on the proton decay rate.}
\begin{equation*}
    m^{*}_{u} \simeq  m_{P}/3.
\end{equation*}
It can be checked by explicit computation that with equal scalar masses and Yukawa couplings all set to unity,
the existing limits on $p\rightarrow \overline{\nu}\overline{\nu}e^{+}/\mu^{+}$ set a lower bound of about 1 TeV on the masses of the new scalar fields, well below the GUT scale and within reach of the LHC.
This lower bound, however, may be further weakened by the constraints from the baryon number conserving probes of the new scalars, which will be discussed in the following subsection.

\subsection{Baryon Number Conserving Constraints}
\label{sec:BNC_constraints}
Let us now discuss the relevant baryon number conserving (BNC) constraints on the new scalar fields.
A novel feature of our model is that there are some processes for which the leptoquark and the dilepton contribute together at the amplitude level. For example, measurements of the anomalous magnetic moment of leptons are sensitive to contributions from both of them. However, due to the maximally chiral nature of the Yukawas in our model, these are suppressed by $m^{2}_{l}/(m_{\chis{j}}m_{\chis{k}})$ \cite{Bigaran:2020jil}, with $j,k=LL,LQ$, hence for $\mathcal{O}$(TeV) masses, these constraints are easily satisfied. A much stronger probe comes from their contributions to the decay of a muon into an electron and a photon, $\mu \rightarrow e\gamma$. This process has a predicted SM rate that renders it unobservable for all practical purposes and can therefore serve as a background-free and sensitive probe of new physics. Expressions for the $\mu\rightarrow e\gamma$ decay rate from theories with only one leptoquark or only one dilepton singlet have been computed in \cite{Bigaran:2020jil,Dorsner:2016wpm,McLaughlin:1999rr}. Making the simplifying assumption that $m_{\rm SM}/m_{\chi} \ll$ 1, where $m_{\rm SM}$ is the mass of \emph{any} SM fermion and $m_{\chi}$ is the mass of the dilepton singlet or leptoquark, the branching ratio for this decay mode can be written as
\begin{widetext}
\begin{equation}
    \text{Br}_{\mu \rightarrow e\gamma} = 12\pi\alpha_{EM}\times \left(\frac{1}{m^{2}_{\chi_{_{LQ}}}G_{F}}\right)^{2}\times \left[N^{2}_{c}\left(\frac{Q_{LQ}}{4}-\frac{1}{6}\right)\sum^{3}_{j=1}y^{*\mu j}_{_{LQ}}y^{ej}_{_{LQ}} + \frac{m^{2}_{_{\chis{LQ}}}}{m^{2}_{_{\chis{LL}}}}\left(\frac{Q_{LL}}{4}-\frac{1}{6}\right)y^{*e\tau}_{_{LL}}y^{\mu \tau}_{_{LL}} \right]^{2},
\end{equation}
\end{widetext}
where $Q_{LQ} = -1/3$ and $Q_{LL} = +1$ are the electromagnetic charges of the leptoquark and the dilepton respectively, and $N_{c} = 3$ is the colour factor. From this expression, it can be checked that if the Yukawa couplings of the leptoquark and/or the dilepton to muons and electrons are set to unity, the combined limit of 
\begin{equation*}
   \text{Br}_{\mu\rightarrow e\gamma} < 3\times10^{-13} 
\end{equation*}
reported by the MEG \cite{MEG:2016leq} and MEG II \cite{MEGII:2023ltw} experiments, requires these two scalars to be heavier than $\mathcal{O}$(10) TeV. 
We note that the lowest order diagram for the $\mu\rightarrow e\gamma$ process via the dilepton singlet must involve a $\tau$ neutrino that couples to both $\mu$ and $e$. To accommodate this stringent constraint, we set $y^{e\tau}_{_{LL}} = 10^{-3}$ for the benchmark used for the collider simulation in the next section,  
allowing for $m_{\chis{LL}}\sim\mathcal{O}$(TeV). 
For the leptoquark, we assume that they couple diagonally across the lepton and quark families, except for the top quark for which we allow, in addition to its coupling to the $\tau$ lepton, an $\mathcal{O}$(1) coupling to the muons. This choice of parameters prohibits lepton flavour violation at tree level and satisfies the above mentioned limit on the ${\mu\rightarrow e\gamma}$ branching fraction.\footnote{We note the rather uninteresting possibility that fine tuning between the sum of leptoquark Yukawas and that of the dilepton singlet to the $\tau$ generation in the limit of nearly degenerate scalar masses can lead to cancellations that provide a further avenue to avoid the stringent constraints from observed limits on the rate for the $\mu \rightarrow e \gamma$ process. However, we choose not to chase that particular ambulance in this work.} This is also a common choice in the LHC searches for leptoquark \cite{ATLAS:2021oiz,CMS:2022nty,ATLAS:2020xov}.\\ 

Next, we discuss the remaining constraints on the leptoquark $\chis{LQ}$.
The tree-level interactions between the leptoquark and the fermions produce dimension six low energy effective operators such as 
\begin{equation*}
    \mathcal{L}_{\text{eff}} = \frac{c_{ijkm}}{m^{2}_{\chi_{_{LQ}}}}\left(\overline{u}^{i}_{L}\gamma^{\mu}d^{j}_{L}\right)\left(\overline{l}^{k}_{L}\gamma_{\mu} \nu^{m}_{L}\right),
\end{equation*}
that contribute to leptonic and semi-leptonic meson decays as well as $\tau$ lepton decays. In the expression above, $i,j$ and $k,m$ label quark and lepton flavours respectively, and $c_{ijkm}$ is the Wilson coefficient that results from integrating out the leptoquark. There are stringent limits on these effective operators from rare meson decays, an up to date review of which can be found in \cite{ParticleDataGroupLeptoquarks:2024cfk}, which we summarize in~\cref{tab:leptoquarklimits}.
\begin{table}[h]
    \centering
 \begin{tabular}{|c|c|}
    \hline
    \textbf{Yukawa Coupling} &  \textbf{Limits}  \\
   \hline
   $ y^{ue}_{LQ} $ & \tiny{$y^{ue}_{LQ} < 0.3 \  (m_{\chi_{LQ}}/\text{TeV})$} \\ 
   \hline
   $y^{c\mu}_{LQ}$ & \tiny{$y^{c\mu}_{LQ}  < 0.8 \  (m_{\chi_{LQ}}/\text{TeV})$}
   \\
   \hline
   
   \end{tabular}
   \caption{Summary of existing constraints on the leptoquark reproduced from \cite{ParticleDataGroupLeptoquarks:2024cfk}. Cross-generational couplings of light quarks to light leptons are set to zero to avoid lepton flavour violation constraints such as the $\mu \rightarrow e \gamma$ process discussed previously. Top quark couplings to the muon and $\tau$ leptons are set to unity while that to the electrons is set to zero for consistency with $\mu \rightarrow e\gamma$ limits. See the text for more details.}\label{tab:leptoquarklimits}
\end{table}
For the remainder of this work, we focus on those with preferential couplings to the third generation quarks and/or leptons, particularly the top quark. In contrast to the leptoquark couplings to the lighter quarks, those to the top quark can only be efficiently probed through their decay to top quarks at the LHC. Such leptoquarks can be produced in pairs via gluon fusion, a channel that is independent of any Yukawa couplings (the production part) and has been searched for by both the ATLAS and CMS experiments \cite{ATLAS:2021oiz,CMS:2022nty}. In particular, the ATLAS experiment set a limit of 1.48 TeV, 1.47 TeV, and 1.43 TeV for scalar leptoquarks decaying to a top quark and an electron, a top quark and a muon, and a top quark and a $\tau$ lepton respectively \cite{ATLAS:2020xov,ATLAS:2021oiz} with unit Yukawa couplings. \\

\begin{table}[h]
    \centering
 \begin{tabular}{|c|c|}
    \hline
    \textbf{Yukawa Coupling} &  \textbf{Limits}  \\
   \hline
   $ y^{e\mu}_{LL} $ & \tiny{$y^{e\mu}_{LL} < 0.08 \  (m_{\chi_{LL}}/\text{TeV})$} \\ 
   \hline
   $y^{\mu\tau}_{LL}$ & \tiny{$ \sqrt{\vert \vert y^{\mu\tau}_{LL}\vert^{2} - \vert y^{e\mu}_{LL}\vert^{2}\vert} < 0.2 \  (m_{\chi_{LL}}/\text{TeV})$}
   \\
   \hline
   $y^{e\tau}_{LL}$ & \tiny{$y^{e\tau}_{LL}$ < 0.001. See $\mu\rightarrow e\gamma$ for details.} \\   
   \hline  
    
   \end{tabular}
   \caption{Summary of existing constraints on the dilepton singlet from \cite{Herrero-Garcia:2014hfa}. See the text for more details.}\label{tab:dileptonsingletlimits}
\end{table}

Then, we have the dilepton singlet $\chis{LL}$. 
This scalar field \emph{necessarily} violates lepton flavor at tree level, making it particularly sensitive to experiments searching for lepton flavor violation or testing lepton flavour universality, particularly in $\tau$ decays. 
Using measurements of leptonic vs. hadronic universality in meson decays and $e/\mu$ universality in $\tau$ decays reported in \cite{Pich:2013lsa}, the authors of \cite{Herrero-Garcia:2014hfa} obtained limits on $y^{e\mu}_{LL}$ and $y^{\mu\tau}_{LL}$, respectively.
We list these limits and the limits from $\mu\to e\gamma$ in~\cref{tab:dileptonsingletlimits} for convenience.
For the benchmark used for collider simulation, we set $y^{\mu\tau}_{_{LL}} = 0.325$ to saturate the lepton universality limit, while also satisfying $\mu\rightarrow e\gamma$. \\

Finally, we discuss the constraints on $\chis{uu}$. There are 3 Yukawa couplings, $y_{uu}^{uc}$, $y^{ut}_{uu}$ and $y^{ct}_{uu}$. For the latter two, there are constraints from the $\Delta F = 2$ processes from flavour physics, particularly $D_{0} \rightarrow \overline{D}_{0}$ mixing. As explained in \cite{Giudice:2011ak}, such processes occur at the one-loop level and, due to the chirality and anti-symmetric nature of the couplings, necessarily involve a box diagram of two diquarks and two top quarks. The authors of \cite{Giudice:2011ak,Pascual-Dias:2020hxo} find particularly tight constraints on the up-top and charm-top Yukawas, 
    \begin{equation}\label{eq:yutCoupling}
         \vert y^{ut}_{uu}y^{ct}_{uu}\vert \times \frac{m_{\chi_{_{uu}}}}{\text{TeV}} \leq 1.5\times 10^{-2}.
    \end{equation}
These bounds do not affect the main production channel for diquarks at the LHC, the $s$-channel process $u \ c \rightarrow \chis{uu}^\dagger$, but favor dijet final states instead of top and jet ones. The former has been searched for by both the ATLAS and CMS experiments~\cite{ATLAS-CONF-2019-007,CMS:2019gwf}, providing strong constraints on $\chi_{_{uu}}$ that are relevant to our model.
These constraints, however, would be weakened by the extra BNV decay mode $\chis{uu}^\dagger\to\chis{LL}^\dagger \chis{LQ}$ if it is kinematically allowed.
We will apply the dijet constraints to our studies of collider phenomenology to be discussed in the next section. For diquark masses above 3 TeV, Yukawa couplings $\vert y^{uc}_{uu}\vert $ = 0.25 are allowed by existing dijet searches, and we adopt this benchmark for the remainder of this work. \\

\subsection{Summary}
\begin{figure}[b!]
\centering
\includegraphics[width=0.48\textwidth]{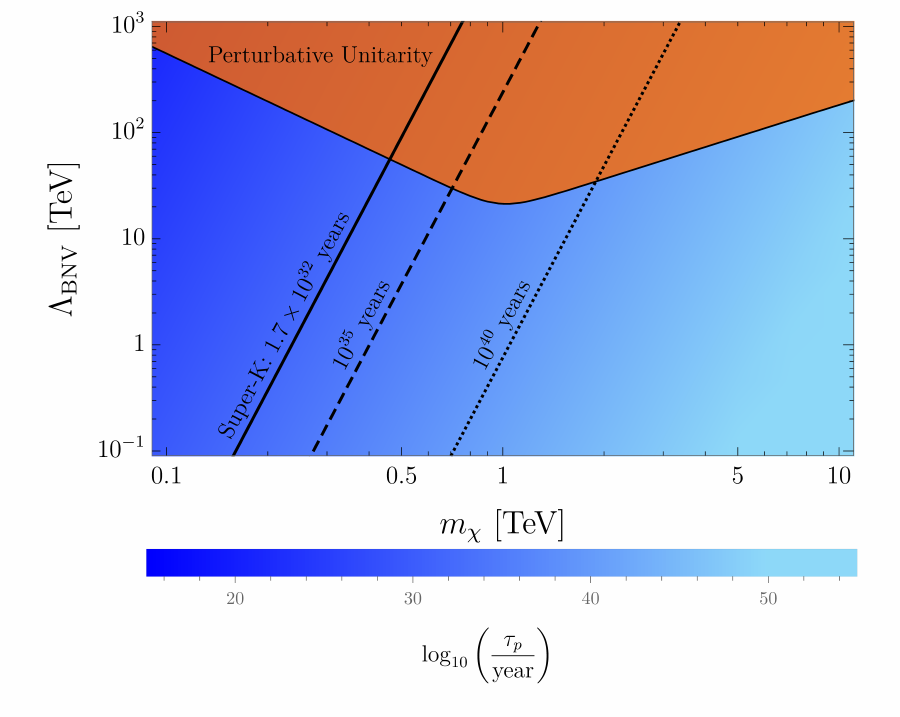}
\caption{Density plot (blue) and constraint (solid black line) on the trilinear coupling, $\Lambda_{\text{BNV}}$, from the lower limit on proton lifetime assuming a $p \rightarrow \overline{\nu}\overline{\nu}e^{+}/\mu^{+}$ decay mode reported by the Super-K \cite{Super-Kamiokande:2014pqx} experiment as well as a $10^{35}$ year lifetime contour potentially achievable at Hyper-K \cite{Hyper-Kamiokande:2018ofw} and $10^{40}$ year one considered ``unrealistic''. We show only the positron mode but since the measured lifetimes from the anti-muon are very similar in value the contours overlap and are not shown in the Figure. For simplicity, in this plot we fix the three scalars to have equal mass and set all relevant Yukawa couplings to values which saturate the bounds in Tables~\eqref{tab:dileptonsingletlimits} and~\cref{tab:leptoquarklimits}. Perturbative unitary constraints, evaluated at $s$ = 4.5$m^{2}_{\chi}$ and described more detail in~\cref{sec:PerturbativeUnitarity}, are shown in orange.}
\label{fig:PDLimits}
\end{figure}
To demonstrate the combined effects of the constraints from proton decay and BNC searches, we show a density plot of the predicted proton lifetime in Figure~\eqref{fig:PDLimits}, which is made with the parameters consistent with the BNC constraints. Collectively, these enhance the proton lifetime by a factor of $\mathcal{O}(10^{6})$ compared to what one would naively expect assuming unit couplings.
For simplicity, we have set all three scalar masses to be equal to $m_{\chi}$.
In solid black we show the lower limit of $1.7\times10^{32}$ years reported by the Super-Kamiokande experiment \cite{Super-Kamiokande:2014pqx} experiment for the $p \rightarrow \overline{\nu}\overline{\nu}e^{+}$ mode and in orange we show perturbative unitarity constraints on $\Lambda_{\text{BNV}}$, described in Appendix~\eqref{sec:PerturbativeUnitarity}. As can be seen from our density plot, for values of scalar mass exceeding 1 TeV, the predicted proton lifetime greatly exceeds the reach of even future experiments like Hyper-Kamiokande \cite{Hyper-Kamiokande:2018ofw} but are consistent with cross-sections for BNV processes that can be probed by the LHC and HL-LHC as will be discussed in the next section.\\

\section{Baryon Number Violation at the LHC and HL-LHC}\label{sec:collider_pheno}

The survey of existing constraints just presented raises the interesting possibility that our model can be probed at the LHC and, subsequently, HL-LHC. LHC searches for BNV, particularly in the top sector \cite{Dong:2011rh}, have been conducted in the past --- see, for example, \cite{CMS:2024dzv}. However, we have not been able to identify a previous search with signatures that are optimized to our model, due in part to its specificity. We therefore propose a dedicated search that offers the opportunity to test it both with current data at the LHC and at the upcoming HL-LHC. \\ 

 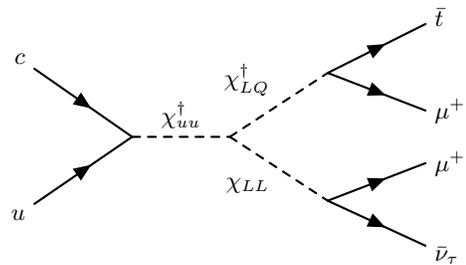
\begin{figure}[t!]
\centering
\begin{tikzpicture}
  \begin{feynman}
    \vertex (uu);
    \vertex [above left=0.85cm and \mylength of uu] (c) {$c$};
    \vertex [below left=0.85cm and \mylength of uu](u) {$u$};
    \vertex [right=\mylength of uu] (bnv);
    \vertex [above right=0.85cm and \mylength of bnv] (LQ);
    \vertex [below right=0.85cm and \mylength of bnv] (LL);
    \vertex [above right=0.5cm and \mylength of LQ] (t) {$\Bar{t}$};
    \vertex [below right=0.5cm and \mylength of LQ, label=right:$\mu^{+}$] (mu);
    \vertex [above right=0.5cm and \mylength of LL, label=right:$\mu^{+}$] (mu2);
    \vertex [below right=0.5cm and \mylength of LL] (nu) {$\Bar{\nu}_{\tau}$};

    \diagram* {
    (c) -- [fermion](uu) --[scalar, edge label=$\chi_{uu}^\dagger$](bnv) -- [scalar, edge label=$\chi_{LQ}^\dagger$](LQ) --[fermion] (t);
    (LQ)-- [fermion](mu);
    (bnv)-- [scalar, edge label'=$\chi_{LL}$](LL) -- [fermion](mu2);
    (LL) -- [fermion](nu);
    (u) -- [fermion](uu);
    };
  \end{feynman}
\end{tikzpicture}
\caption{Parton level description of the distinctive hadron collider signature we propose for our model. An up and a charm quark from the protons annihilate to produce a diquark, a process which benefits from an $s$ channel enhancement. It then decays to a leptoquark and a dilepton singlet. These in turn decay to an anti-top quark and an anti-muon, and an anti-muon and an anti-neutrino. The anti-top in the final state makes it possible to identify the charge --- and hence baryon number of the outgoing particle and establish that baryon number violation by one unit, $\Delta \text{B} = -1$, has occurred. However, due to the anti-neutrino in the final state, it is not possible to identify whether lepton number has bee violated by one, $\Delta \text{L} = -1$, or three, $\Delta \text{L} = -3$, units. }
\label{fig:ColliderSignatureFinal}
\end{figure}

Our proposal is based, at the parton level, on the Feynman diagram shown in Figure~\eqref{fig:ColliderSignatureFinal}. The diquark is produced on-shell via the annihilation process  $u c \rightarrow \chi^{\dagger}_{_{uu}}$ and decays to the leptoquark $\chis{LQ}^{\dagger}$ and the dilepton singlet $\chis{LL}$. These in turn decay to an anti-top quark and an anti-muon $\bar{t} \mu^+$, and an anti-muon and an anti-neutrino $\mu^+\bar{\nu}_\tau$: 
\begin{equation}
    uc \rightarrow \chis{uu}^\dagger \rightarrow \chis{LQ}^{\dagger} \chis{LL}  \rightarrow \bar{t}\mu^+ \mu^+\bar{\nu}_{\tau}.
    \label{eq:LHCsignal}
\end{equation}
A similar process with two anti-$\tau$ leptons instead of two anti-muons is possible and, taking existing constraints into account, has a similar cross-section to the process we consider. The CP conjugate of these processes, while being potentially within reach of the LHC, are one order of magnitude suppressed compared to these ones due to the lower anti-quark content of the proton and will not be considered further. We also note in passing that even though the LHC is insensitive to neutrino flavours, due to existing constraints described in the previous section, the dilepton singlet decay to $e^{+}\overline{\nu}_{\mu}$ or $\mu^{+}\overline{\nu}_{e}$ is more suppressed. \\

Having an anti-top quark in the final state makes it easier to reconstruct the baryon number of the final states and ascertain that BNV by one unit occurred, $\Delta \text{B} = -1$. Particle and charge identification of the anti-top quark requires leptonic decay of the associated $W$-boson. While leptonic decay adds another missing particle to the final state, the charged lepton and $b$-jet separation will be small since the anti-top quark should be boosted~\cite{Rehermann:2010vq}. Full kinematic reconstruction of the anti-top quark is then not necessary, as only the particle and charge identification is needed to reconstruct the baryon number. If full reconstruction of the anti-top quark is necessary, then hadronic decay of the $W$-boson can be used at the expense of reconstructing the baryon number of the final state. We leave the full details of the requirements of anti-top quark reconstruction for a proper experimental analysis of our model. On the other hand, due to the anti-neutrino in the final state, lepton number cannot be fully reconstructed and its violation is determined to be either $\Delta \text{L} =-1$ or $\Delta \text{L} =-3$. Interestingly, one can reconstruct the lepton number of the dilepton singlet from the 
\begin{equation*}
    p p \rightarrow W^{*}\rightarrow e^{-}\mu^{-}\chi_{_{LL}},\chi_{_{LL}} \rightarrow \mu^{+} \overline{\nu}_{\tau},
\end{equation*}
process and its CP conjugate version by making use of the fact that the original $W$ vertex must have zero lepton number. This can help identify the correct $\Delta L = -3$ violation for the BNV process, but, unfortunately, we find that existing constraints predict a cross-section for this process that is too small to yield a statistically significant number of events at the HL-LHC. \\

\begin{table}[b!]
    \centering
 \begin{tabular}
 {|c |c |c |c |}
 \hline
 \textbf{$\sigma \times \text{BR} $ (pb)} & \textbf{Signal BP1} & \textbf{Signal BP2}  & \textbf{$t\overline{t}W$} \\
 \hline
 Before cuts& 8.3 $\times \ 10^{-4}$ & 3.6 $\times \ 10^{-5}$ & 2.7 $\times \ 10^{-3}$ \\
 \hline
 Basic cuts& 5.5 $\times \ 10^{-4}$ & 2.5 $\times \ 10^{-5}$ & 5.2 $\times 10^{-6}$ \\
 \hline
 \end{tabular}
 \caption{Cut flow of the signal at each benchmark point (BP) and dominant SM background $t\bar{t}W$. The cuts applied are given in~\cref{eq:LHCCuts}.}
 \label{tab:cutflow}
\end{table}

The signature we propose shares many similarities with previous same-sign dilepton searches performed by both the ATLAS \cite{ATLAS:2011izm} and CMS \cite{CMS:2016mku} experiments. While these same-sign dilepton searches could in principle probe our model, we expect that requiring the anti-top quark will make existing searches less efficient at probing the model parameter space, since the existing searches require only 2 or more jets without the use of top-tagging. We therefore suggest a dedicated search strategy, using existing same-sign dilepton searches as inspiration for kinematic cuts and background determination. We adopt the main sources of background events considered in these works, listed here: 
\begin{enumerate}
    \item Non-prompt leptons which can be suppressed by a large transverse momentum cut, $p_{T} > $ 25 GeV on each lepton in the di-lepton pair together with a rapidity cut $\vert \eta_{\ell} \vert < 2.5$.
    \item The dominant SM process mimicking our signal is $p p \rightarrow t\overline{t}W^{+}$, with the top and $W$ decaying to two anti-muons. This process comes with an additional $b$ jet not present in our signal and can be further suppressed using a jet veto that we do not consider here. 
    \item Another process that can potentially contribute is $p p \rightarrow W Z j, Z \rightarrow \mu^{+} \mu^{-}$ when the jet is misidentified as an anti-top quark and the outgoing muon is not detected and therefore considered as missing energy. We have found that, assuming a conservative jet and lepton misidentification\footnote{More appropriately non-identification for the lepton.} probability of 10$\%$ each, this background is sub-dominant. 
    \item The most important cut in terms of background suppression is on the missing energy. We set $E^{\text{miss}}_{T} > 500$ GeV, which is higher than that considered in \cite{CMS:2016mku}. Our choice of higher $E^{\text{miss}}_{T}$ is motivated by the fact that the missing energy in our process is carried by a single outgoing anti-neutrino produced from the two-body decay of a TeV mass dilepton singlet and is therefore naturally peaked at very high energies compared to SM neutrinos. The latter are produced mainly from $W$ boson decays and therefore peak at lower energies, making this $E^{\text{miss}}_{T}$ cut particularly effective at suppressing SM backgrounds while not affecting signal events in a significant way.
\end{enumerate} 

\begin{figure}[t!]
\centering
\subfloat[LHC Run 2: $\sqrt{s} = 13$ TeV, $\mathcal{L} = 139 \text{ fb}^{-1}$]{
    \centering
    \includegraphics[width=0.38\textwidth]{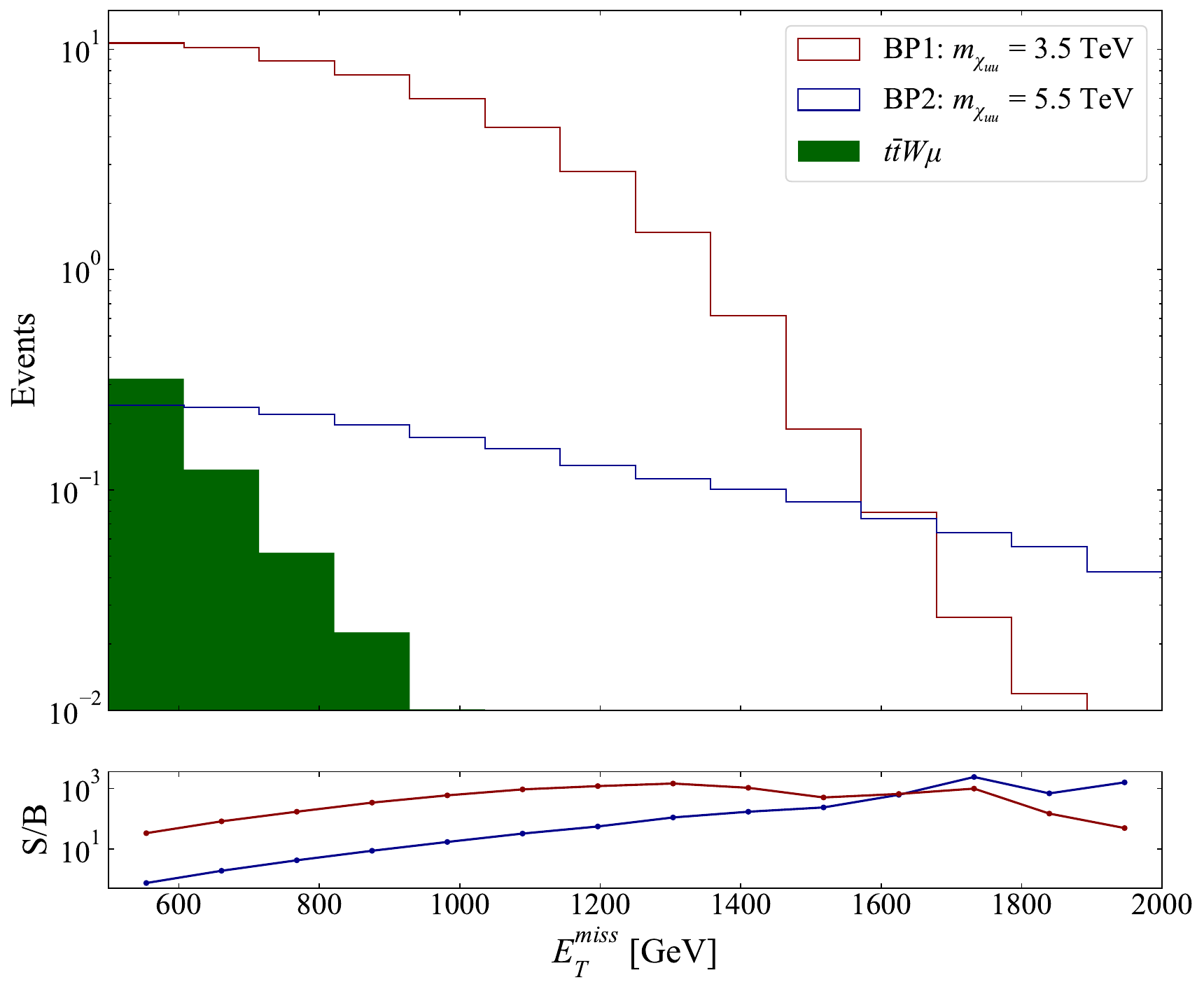}
}

\subfloat[HL-LHC: $\sqrt{s} = 14$ TeV, $\mathcal{L} = 3000 \text{ fb}^{-1}$]{
    \centering
    \includegraphics[width=0.38\textwidth]{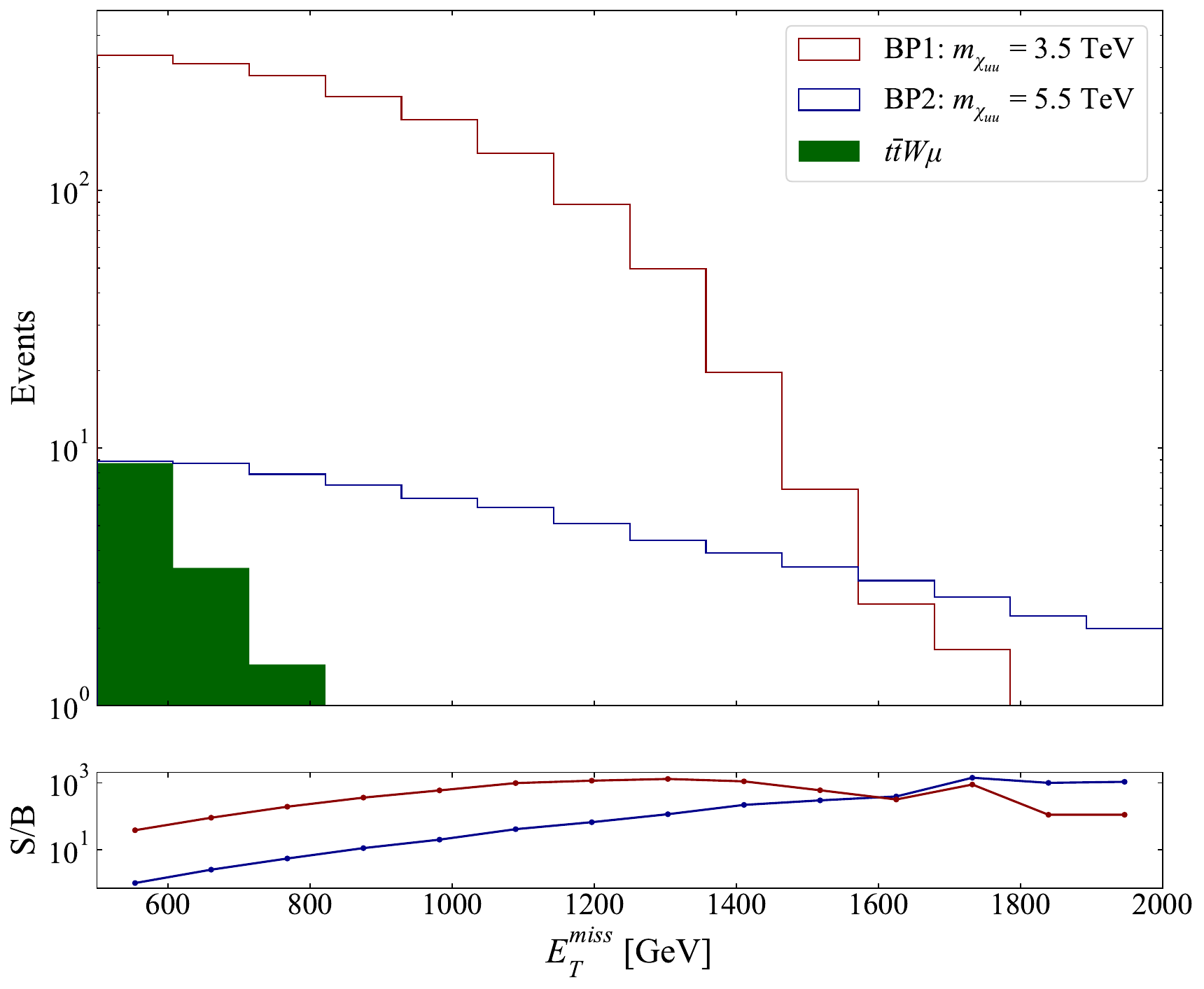}
}
\caption{Missing transverse energy distributions for signal and background events for the current LHC and HL-LHC cases. The signal to background ratios for each bin are given in the subplots.}
\label{fig:BNVDistributionPlots}
\end{figure}

We define the basic selection cuts as follows:
\begin{equation}
 p_{T,\ell} > 25 \ \text{GeV}, \ \vert \eta_{\ell,j,b,t} \vert < 2.5, \ E^{\text{miss}}_{T} > 500 \ \text{GeV}.
\label{eq:LHCCuts}
 \end{equation}
As an illustrative example, we consider two benchmark points, with diquark masses $m_{\chi_{_{uu}}} = 3.5$ TeV (BP1) and $m_{\chi_{_{uu}}} = 5.5$ TeV (BP2). The remaining parameters are fixed, with the following Yukawa matrices chosen to saturate the BNC bounds:
\begin{align}
    y_{uu} &= \begin{pmatrix}
                0 & 0.25 & 0.12 \\
                -0.25  & 0 & 0.12 \\
                -0.12  & -0.12 & 0
    \end{pmatrix},\, \nonumber \\ 
    y_{LL} &= \begin{pmatrix}
                0 & 0.125 & 10^{-3} \\
                -0.125 & 0 & 0.325 \\
                -10^{-3} & -0.325 & 0
    \end{pmatrix},\, \\ 
    y_{LQ} &= \begin{pmatrix}
                0.397 & 0 & 0 \\
                0 & 0.949 & 1 \\
                0 & 1 & 1
    \end{pmatrix}. \nonumber
\end{align}
We conventionally pick the upper triangular entries of the anti-symmetric matrices to be positive and set the scalar masses to be $m_{\chis{LQ}}=m_{\chis{LL}}=$ 1.5 TeV and $\Lambda_{\text{BNV}}$ = 10 TeV. We simulate the process shown in Figure~\eqref{fig:ColliderSignatureFinal} using \texttt{MadGraph5\_aMC@NLO}~\cite{Alwall:2014hca} for the LHC and for its HL upgrade. The cross sections for the signal at two benchmark points,  at $\sqrt{s} = 14$ TeV as well as the main background before and after our chosen cuts are shown in ~\cref{tab:cutflow}. \\

\begin{figure}[b!]
\includegraphics[width=0.45\textwidth]{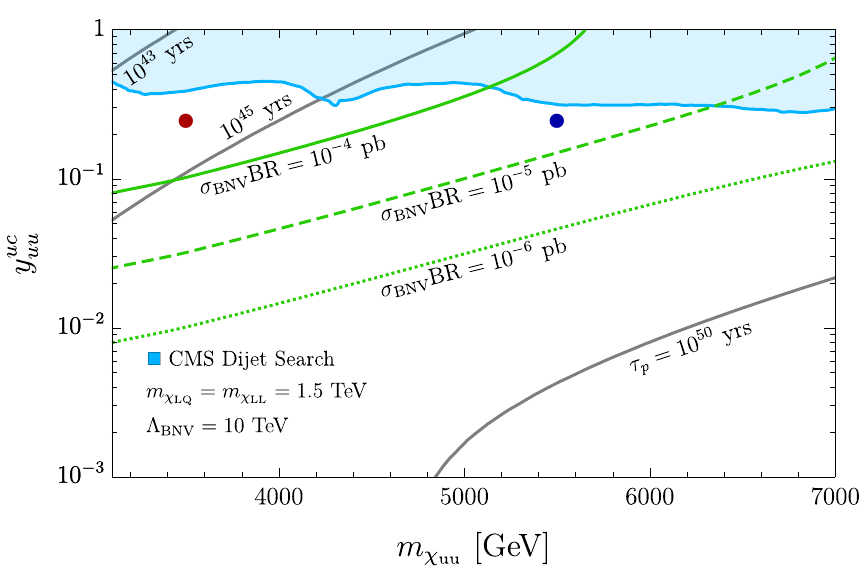}
\centering
\caption{Parameter space showing existing constraints (blue), proton lifetime contours (dark gray contours), and BNV cross sections at $\sqrt{s} = 14$ TeV observable at the HL-LHC (green) while varying the scalar diquark mass and Yukawa coupling to up quark and charm quark. The shaded blue region is ruled out by existing dijet searches at the LHC~\cite{ATLAS-CONF-2019-007,CMS:2019gwf}. The proton lifetimes in this region span from $\tau_p \gtrsim 10^{43}$ years, well above the current Super-K bound ($\sim 10^{32}$ years) and projected Hyper-K bound ($\sim 10^{35}$ years). The green contours show the BNV cross section times branching ratio for the final state of ~\cref{eq:LHCsignal} after applying the cuts of ~\cref{eq:LHCCuts}. They correspond to $10^{-4}$ pb (solid), $10^{-5}$ pb (dashed), and $10^{-6}$ pb (dotted), accessible at the (HL-)LHC. The solid contour corresponds to $\mathcal{O}(300)$ events with HL-LHC luminosity $\mathcal{L}=3 \times 10^6$ pb$^{-1}$. The dark red and dark blue circles represent the benchmark points shown in~\cref{fig:BNVDistributionPlots}.}
\label{fig:ParamSpace}
\end{figure}

In ~\cref{fig:BNVDistributionPlots} we show the resulting event distributions for the missing transverse energies in our chosen final states. We find that for $m_{\chi_{uu}} \in[3.5 \text{ TeV}, 5.5 \text{ TeV}]$, the predicted number of signal events is significantly in excess of the background events coming mainly from $t\overline{t}W$, highlighting the potential of the LHC and HL-LHC to probe this region of parameter space. Therefore, an ideal target search for our model even using current LHC data is to look for an excess of missing transverse energy in the $\bar{t}\mu^+\mu^+$ final state. The reach of the HL-LHC in probing our model (green) is shown in ~\cref{fig:ParamSpace} for our benchmark discussed previously, but varying the scalar diquark mass and Yukawa coupling to up quark and charm quark. We also show, on the same figure, existing dijet constraints (blue) and several contours for the proton lifetime that are all well above $10^{35}$ years, to illustrate that to probe our model, proton decay experiments are sub-optimal compared to collider experiments.

\newcommand\minisec[1]{\emph{\textbf{#1}}.}
\newcommand{\mr}{m_{\Phi_r}}
\section{Baryogenesis}\label{sec:baryogenesis}
Baryogenesis within our BNV model can be realized through a mechanism similar to the post-sphaleron baryogenesis mechanism developed in \cite{Babu:2006xc,Babu:2013yca}. Here we consider the scenario where $\Lambda_{\rm BNV}$ is set by the vacuum expectation value of a complex scalar field $\Phi$ as discussed in \cref{sec:Model}.
After the spontaneous symmetry breaking of B and L, the real part of $\Phi$ --- denoted as $\Phi_r$ --- acquires a mass $m_{\Phi_r}$ which can be $\mathcal{O}(100)\,$GeV.
The decay of $\Phi_r$ can generate baryon asymmetry if the Sakharov's conditions \cite{Sakharov:1967dj} are satisfied:
\begin{enumerate}
    \item{ Baryon number violation}
    \item{ Out of thermal equilibrium}
    \item{ C, CP violation}
\end{enumerate}
The first condition is apparently satisfied since the only decay channel of $\Phi_r$ to the SM fermions is mediated by the BNV operator of our model as shown in \cref{fig:cp-asymmetry}.
The second and third conditions can also be naturally satisfied as discussed below.\\

The out-of-thermal-equilibrium condition involves comparing the decay rate of $\Phi_r$ to the expansion rate of the universe $H$.
At a temperature below $\mr$, the six-body decay rate for $\Phi_r\to 3L+Q+2u_R$ mediated by the new scalar fields is independent of the temperature and is estimated to be \cite{Babu:2013yca}: 
\begin{align}
    \Gamma_{r}=&\Gamma(\Phi_r\to 6f) + \Gamma(\Phi_r\to 6\Bar{f})\nonumber\\
    \simeq&\frac{2\cdot 6 P}{\pi^9\cdot 2^{25}\cdot 45}|\lambda|^2 \,\text{Tr}(\ychi{LL}^\dagger \ychi{LL})\,\text{Tr}(\ychi{LQ}^\dagger \ychi{LQ})\,\text{Tr}(\ychi{uu}^\dagger \ychi{uu})\nonumber\\
    &\cdot\left(\frac{\mr^{13}}{m_{\chis{LL}}^{4}m_{\chis{LQ}}^{4}m_{\chis{uu}}^{4}}\right),
\end{align}
where $\lambda$ is a dimensionless coupling constant that couples $\Phi$ to the three $\chi$ fields, and $6$ is the color factor.
$P$ is the phase space factor evaluated to be $P\sim 10^{-4}$ in \cite{Babu:2013yca}.
For the following estimations, we use the same Yukawa couplings and $m_\chi$ as the benchmark chosen in \cref{sec:collider_pheno}, and set $\lambda=1$.
For $\mr\sim\mathcal{O}(100)\,$GeV and $m_\chi\sim \mathcal{O}$(TeV), $\Gamma_r$ is much lower than the Hubble rate $H(T)=1.66g_{\ast}^{1/2}T^2/M_{Pl}$ at the temperature $T=\mr$.
Hence, the decay of $\Phi_r$ is negligible until the temperature drops to $T_d$ at which $\Gamma_r= H(T_d)$.
When it finally starts to decay at around $T_d$, $\Phi_r$ is already far out of equilibrium, satisfying the out-of-equilibrium condition of baryogenesis.
We find that for $\mr<1\,$TeV, $T_d\lesssim 0.9\,$GeV is well below the electroweak scale $T_{\rm EW}\approx 100\,$GeV, hence the electroweak sphalerons have been deactivated at the time of $\Phi_r$ decay and would not wash out the generated baryon asymmetry.
Requiring the decay to occur prior to the QCD phase transition: $T_d>\Lambda_{\rm QCD}\approx 200\,$MeV, $\mr$ is constrained above a lower bound at $\mr^{\rm min}\approx 790\,$GeV.
The out-of-equilibrium decay of $\Phi_r$ would produce entropy that would dilute the baryon asymmetry.
The dilution factor is given by $\eta\approx [1 + \mr/(g_\ast T_d)]^{-1}$ \cite{Luty:1992un}, which varies in the range of $0.03\lesssim\eta \lesssim 0.09$ for $\mr^{\rm min}<\mr<1\,$TeV.\\

CP violation can be induced from the interference of the tree and one-loop-level diagrams shown in \cref{fig:cp-asymmetry}.
In this work, we only consider the interference of the tree-level diagram with the $W$-loop correction to the $\chis{LQ}$ vertex as shown in the middle panel of \cref{fig:cp-asymmetry}, 
which has an appealing feature of depending solely on the CP violations in the quark and lepton sectors of the Standard Model\footnote{There is another possible $W$-loop correction to the $\chis{LL}$ vertex, which, however, has real coupling factor in the interference term due to the antisymmetric structure of $\ychi{LL}$. Hence it does not contribute to CP asymmetry.}\footnote{The $W$-loop correction on the wavefunctions of the outgoing particles would also contribute to CP asymmetry. This contribution, however, was found to be subdominant to the vertex correction \cite{Babu:2006xc,Babu:2013yca}, and thus is not considered in the present work.}.
The contribution from the $\chis{LL}$--exchange diagram shown in the right panel of \cref{fig:cp-asymmetry} is a possible CP violation source if considering complex Yukawa couplings, while it is vanishing for the real Yukawa couplings considered in this work.
The CP asymmetry quantified by the ratio of the difference in the decay rates of the CP conjugate processes to the sum of them is found to be:
\begin{widetext}
\begin{align}
     \epsilon_{\rm CP}\equiv \frac{\Gamma(\Phi_r\to 6f)-\Gamma(\Phi_r^\ast\to 6\Bar{f})}{\Gamma(\Phi_r\to 6f)+\Gamma(\Phi_r^\ast\to 6\Bar{f})}
     \simeq \frac{g_W^2}{4\pi}\log\left(1+\frac{s}{m_W^2}\right)\frac{{\rm Im}\,{\rm Tr}[ \ychi{LQ}^\dagger U^\ast \ychi{LQ} V^\dagger + \ychi{LQ}^\dagger U^T \ychi{LQ} V]}{{\rm Tr}[\ychi{LQ}^\dagger \ychi{LQ}]},
     \label{eq:CP_asymmetry}
\end{align}
\end{widetext}
where $\sqrt{s}$ is the total energy of $L_{\gamma k}$ and $Q_{\delta l}$ in their center-of-mass frame, $U\equiv V_{L}^l V_L^{\nu\dagger}$ is the PMNS matrix of the lepton sector \cite{pdg_pmns_matrix,Pontecorvo:1957qd,Maki:1962mu}, and $V\equiv V_L^u V_L^{d\dagger}$ is the CKM matrix of the quark sector \cite{pdg_ckm_matrix,Cabibbo:1963yz,Kobayashi:1973fv}.
The $V^f_{L}$ matrices are defined such that the mass matrix of the fermion of flavor $f$ is diagonalized as $M^f_{\rm diag}=V_L^f Y^f V_R^{f\dagger}(v/\sqrt{2})$.
Note that in the kinematic calculation that leads to \cref{eq:CP_asymmetry}, we ignore the fermion masses which are much smaller than the momentum scale. However, we only take this limit at the very end of our calculations, since the PMNS and CKM matrices should be zero in the massless limit. Including the fermion masses only introduces a small correction of $\mathcal{O}(m_f/s)$ to the kinematic factor.

The PMNS matrix remains loosely constrained, so are the CP asymmetry and the baryon asymmetry $\eta_B\sim\epsilon_{\rm CP}\cdot\eta$ generated from the presented mechanism.
Hence the observed baryon asymmetry $\eta_B^{\rm obs}\sim 10^{-10}$ can be accommodated in a large parameter space of our model at present, and will be directly constrained by the future neutrino precision measurements. 

\begin{widetext}
\begin{figure*}[t!]
\subfloat
{\resizebox{0.25\textwidth}{!}{
    \begin{tikzpicture}
  \begin{feynman}
    \vertex (b);
    \vertex [left=1.cm of b] (i1) {$\Phi_r$};
    \vertex [above right=0.85cm and 1.cm of b] (LL);
    \vertex [above right=0.85cm and 1.cm of LL] (Li) {$\LL{\alpha}{i}$};
    \vertex [right=1.cm of LL] (Lj) {$\LL{\beta}{j}$};
    \vertex [right=1.5cm of b] (LQ);
    \vertex [above right=0.25cm and 1.0cm of LQ, label=right:$\LL{\gamma}{k}$] (Lk);
    \vertex [below right=0.25cm and 1.0cm of LQ, label=right:$\QL{\delta}{l}$] (Ql);
     \vertex [below right=0.85cm and 1.cm of b] (uu);
     \vertex [below right=0.85cm and 1.cm of uu] (un) {$\uR{n}$};
     \vertex [right = 1.cm of uu] (um) {$\uR{m}$};

    \diagram* {
      (i1) -- [scalar] (b) -- [scalar, edge label=$\chis{LL}$] (LL) -- [fermion] (Li),
      (b) -- [scalar, edge label = $\chis{LQ}^\dagger$, near end] (LQ) -- [fermion] (Lk),
      (LL)-- [fermion] (Lj),
      (LQ)-- [fermion] (Ql),
      (b) -- [scalar, edge label' = $\chis{uu}$] (uu) -- [fermion] (um),
      (uu) -- [fermion] (un),
    };
  \end{feynman}
\end{tikzpicture}}}
\subfloat
    {\resizebox{0.3\textwidth}{!}{
    \begin{tikzpicture}
  \begin{feynman}
    \vertex (b);
    \vertex [left=1.cm of b] (i1) {$\Phi_r$};
    \vertex [above right=0.9cm and 1.cm of b] (LL);
    \vertex [above right=0.9cm and 1.cm of LL] (Li) {$\LL{\alpha}{i}$};
    \vertex [right=1.cm of LL] (Lj) {$\LL{\beta}{j}$};
    \vertex [right=1.2cm of b] (LQ);
    \vertex [above right=0.3cm and 1.2cm of LQ] (La);
    \vertex [below right=0.3cm and 1.2cm of LQ] (Qb) ;
    \vertex [right=1cm of La] (Lk) {$\LL{\gamma}{k}$};
    \vertex [right=1cm of Qb] (Ql) {$\QL{\delta}{l}$};
     \vertex [below right=0.9cm and 1.cm of b] (uu);
     \vertex [below right=0.9cm and 1.cm of uu] (un) {$\uR{n}$};
     \vertex [right = 1.cm of uu] (um) {$\uR{m}$};

    \diagram* {
      (i1) -- [scalar] (b) -- [scalar, edge label=$\chis{LL}$] (LL) -- [fermion] (Li),
      (b) -- [scalar, edge label = $\chis{LQ}^\dagger$, near end] (LQ) -- [fermion, edge label=$\LL{\mu}{a}$, near end] (La) -- [fermion] (Lk),
      (LL)-- [fermion] (Lj),
      (LQ)-- [fermion, edge label'=$\QL{\nu}{b}$, near end] (Qb) -- [fermion] (Ql),
      (La)-- [boson, edge label=$W_{\pm}$] (Qb),
      (b) -- [scalar, edge label' = $\chis{uu}$] (uu) -- [fermion] (um),
      (uu) -- [fermion] (un),
    };
  \end{feynman}
\end{tikzpicture}}}
\subfloat
    {\resizebox{0.35\textwidth}{!}{
    \begin{tikzpicture}
  \begin{feynman}
    \vertex (b);
    \vertex [left=1.cm of b] (i1) {$\Phi_r$};
    \vertex [above right=0.85cm and 1.cm of b] (LL);
    \vertex [above right=0.85cm and 1.cm of LL] (Li) {$\LL{\alpha}{i}$};
    \vertex [right=1.5cm of LL] (Loop) ;
    \vertex [above right=0.85cm and 1.cm of Loop] (LL2);
    \vertex [above right=0.85cm and 1.cm of LL2] (Lj) {$\LL{\beta}{j}$};
    \vertex[right=1.5cm of LL2] (Lk) {$\LL{\gamma}{k}$};
    \vertex [right=1.5cm of b] (LQ);
    \vertex [below right=0.25cm and 1.0cm of LQ, label=right:$\QL{\delta}{l}$] (Ql);
     \vertex [below right=0.85cm and 1.cm of b] (uu);
     \vertex [below right=0.85cm and 1.cm of uu] (un) {$\uR{n}$};
     \vertex [right = 1.cm of uu] (um) {$\uR{m}$};

    \diagram* {
      (i1) -- [scalar] (b) -- [scalar, edge label=$\chis{LL}$] (LL) -- [fermion] (Li),
      (b) -- [scalar, edge label = $\chis{LQ}^\dagger$, near end] (LQ) -- [fermion, edge label'=$\LL{\nu}{b}$, near end] (Loop),
      (LL)-- [fermion, edge label=$\LL{\mu}{a}$, near end] (Loop) --[scalar, edge label'=$\chis{LL}$] (LL2) --[fermion] (Lj),
      (LL2)-- [fermion] (Lk),
      (LQ)-- [fermion] (Ql),
      (b) -- [scalar, edge label' = $\chis{uu}$] (uu) -- [fermion] (um),
      (uu) -- [fermion] (un),
    };
  \end{feynman}
\end{tikzpicture}}}
    \centering
    \caption{Tree and one-loop diagrams of $\Phi_r$ decay that would interfere and violate CP symmetry. We use $i,j,k,...$ for the flavour indices, and $\alpha,\beta,\gamma,...$ for the $SU(2)_L$ indices.}
    \label{fig:cp-asymmetry}
\end{figure*}
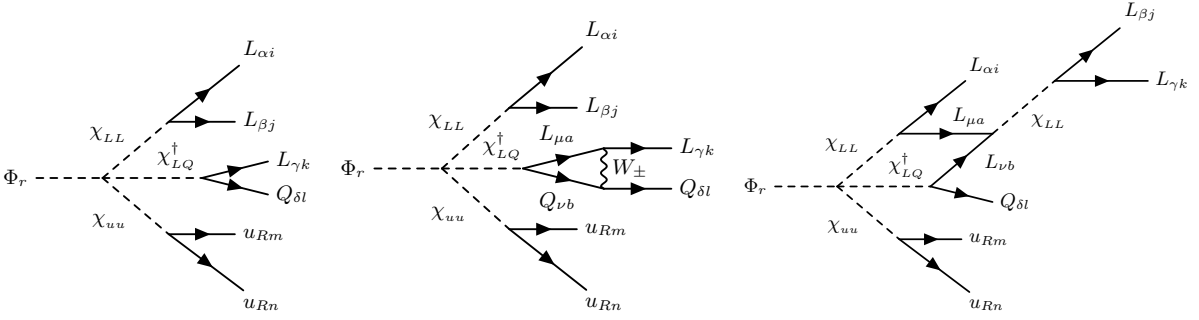
\end{widetext}

\section{Conclusion}\label{sec:conclusion}
In this work, we have introduced a model for baryon number violation by one unit and lepton number violation by three units that adds three new scalars, a diquark, a leptoquark, and a dilepton singlet, to the SM. We showed that limits from proton decay fix the masses of these new scalars to be $\mathcal{O}$(TeV), which opens opportunities to probe this model using both low energy precision and collider experiments. After surveying these, we proposed a dedicated search for this model that can be done with both currently available data at the LHC and upcoming ones at the HL-LHC. We also extended our model to propose a viable scenario for low-scale baryogenesis, below the temperature at which the electroweak sphalerons are active. \\

There are many aspects of our work that merit further investigations. For one, despite violating lepton number, our model is unable to provide a mechanism for neutrino masses. This can be understood from the fact that our model violates lepton number by three units while neutrino masses require it to be violated by two units. A straightforward extension of our model by either an extra Higgs doublet weakly coupled to SM fermions, or a doubly charged scalar coupling to two charged leptons, both also additionally couple to our dilepton singlet would be able to generate neutrino masses radiatively in a manner identical to the Zee-Babu model \cite{Zee:1980ai,Babu:1988ki}. However, in the context of our model, this would also involve new signatures of baryon number violation. For example, the addition of a doubly charged scalar that generates neutrino masses according to the Babu model will also produce a new proton decay mode 
\begin{equation*}
    p \rightarrow e^{+}/\mu^{+} \ \overline{\nu}\overline{\nu} \ e^{-}\mu^{+}/\mu^{-}e^{+}.
\end{equation*}
A systematic study of all the necessary additional interactions required for our model to generate neutrino masses, and their implications for signatures of baryon number violation would be very desirable.
\\ 

Another aspect of our model that should be studied further is whether or not it can be embedded into a non-minimal GUT theory. Since each of the new elementary particles we have introduced appear in grand unification scenarios, it is possible that the B - L/3 symmetry we imposed can arise naturally as an ``accidental'' symmetry of some non-minimal GUT model. In particular, since the Pati-Salam (for the diquark) \cite{Pati:1974yy} and $SU(5)$ (for the leptoquark and dilepton singlet) \cite{Georgi:1974sy} gauge groups can be embedded into $SO(10)$, it should be possible to design a particular symmetry breaking pattern of the latter where the model presented here is an intermediate theory. \\

In terms of phenomenology, it would be interesting to investigate how this model can be further probed, both with near term technological improvements such as developments in charge reconstruction for jet tagging and with longer term experiments such as the planned 
future lepton and hadron colliders~\cite{FCC:2025lpp,CEPCStudyGroup:2018ghi,CEPC-SPPCStudyGroup:2015csa,CLIC:2018fvx,ILC:2013jhg,Accettura:2023ked,InternationalMuonCollider:2025sys}.

Finally, we offer some concluding remarks about baryogenesis.
An interesting feature of our model is that the BNV interaction at low energy and the baryon asymmetry of the universe can be of the same source---a complex scalar singlet that couples to all three $\chi$ fields.
The non-vanishing VEV of this complex singlet determines the BNV coupling strength, while the decay of the real part of it via the BNV vertex could generate baryon asymmetry if it occurs out of equilibrium and breaks CP symmetry.
The out-of-equilibrium condition can be easily satisfied due to the strong suppression of the decay rate from the high dimensionality of the BNV operator, resulting in a decay temperature much lower than the mass of the decaying particle.
If the decaying particle is lighter than 1\,TeV, the decay temperature is well below the electroweak scale, thus preventing the washout by the electroweak sphalerons.
The other condition, CP violation, can be induced purely from the CP violation in the quark and lepton sectors of the Standard Model via the interference of the tree and the $W$-exchanged one-loop diagrams.
This indicates that the baryogenesis mechanism based on our model can be directly constrained by future measurements of the lepton sector CP violation.

\begin{acknowledgments}
We are especially grateful to Kaladi Babu for pointing out perturbative unitary constraints relevant for our model, to Brian Batell for making us aware of constituent quark masses relevant for our proton decay rate calculations, to Tao Han for suggesting the main collider signature for our BNV process, and to Julian Heeck for multiple insightful discussions and, in particular, a careful review of our proton decay rate calculation. We would like to thank Joseph Boudreau, Ayres Freitas, and James Mueller for clarifying issues regarding the unitarity of the CKM and PMNS matrices. We also acknowledge Joshua Berger, Debasish Borah, Kun Cheng, Arnab Dasgupta, Akshay Ghalsasi, and Kuver Sinha for helpful discussions and feedback. The work of DS is partly supported by NSF Grant No.~PHY-2210361 and U.S. DOE Grant DE-SC0010813, by the Maryland Center for Fundamental Physics, and by the Mitchell Institute for Fundamental Physics and Astronomy.
TO is supported by the DOE Office of Science Distinguished Scientist Fellows Award 2022. The work of AB, FB and DL was supported in part by the U.S. Department of Energy under grant No. DE-SC0007914 and in part by Pitt PACC. 
\end{acknowledgments}

\appendix  
\section{Perturbative Unitarity Constraints}\label{sec:PerturbativeUnitarity}
 
An important theoretical constraint on various parameters in the scalar sector, in particular the BNV trilinear coupling, comes from perturbative unitarity. This is determined by, for example, the  $\tiny{\chi_{_{LQ}} \ \chi_{_{LL}} \rightarrow \chi_{_{LQ}} \ \chi_{_{LL}}}$ scattering process, for which there are four diagrams, one involving the mixed quartic $\lambda_{_{LLLQ}},$ and three other ones with the diquark in the mediator corresponding to the $s$, $t$, and $u$ channels. \\ 
  
\begin{figure}[h!]
\includegraphics[width=0.45\textwidth]{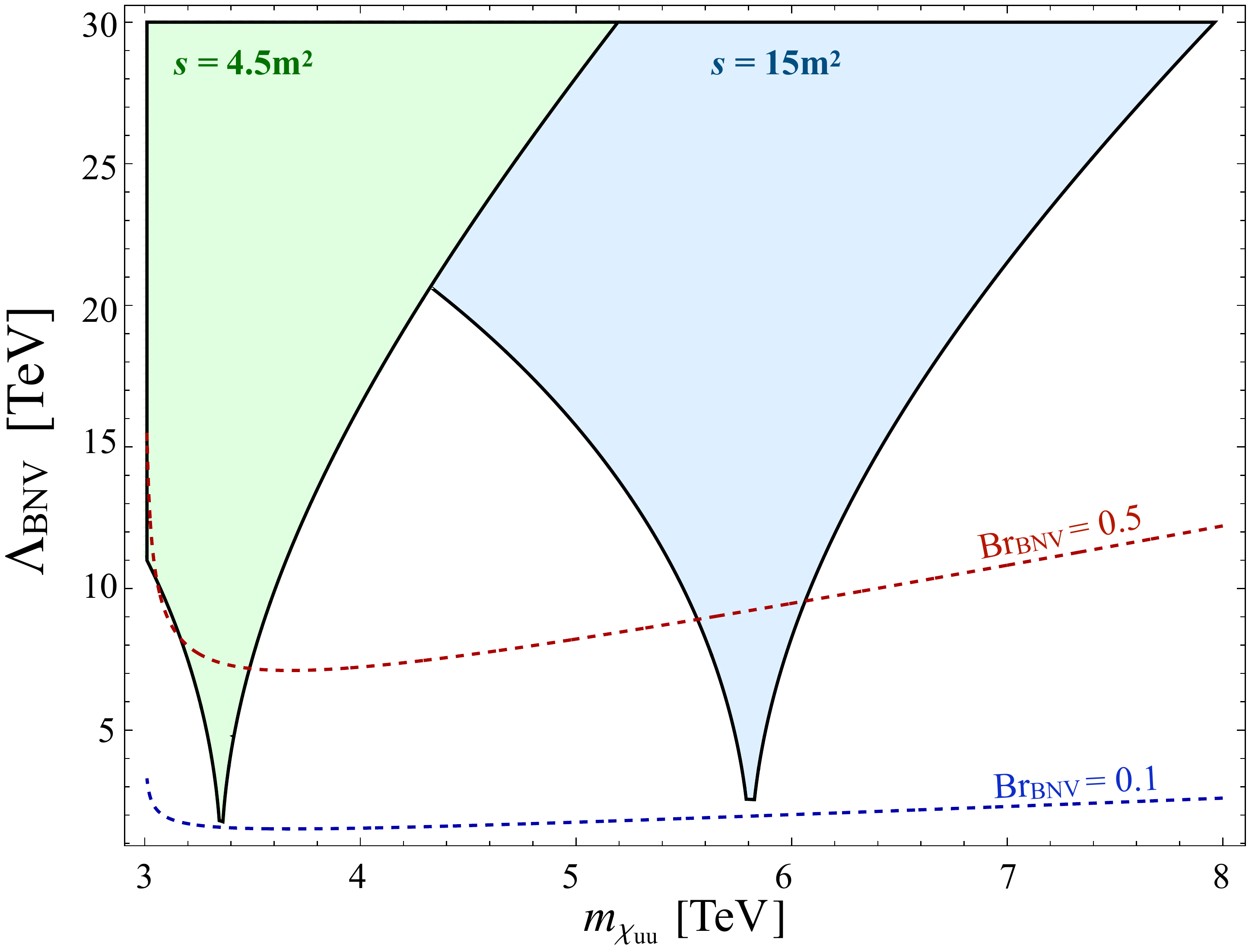}
\centering
\captionsetup{justification=justified,width=0.5\textwidth}
\caption{Perturbative unitarity constraints on $\Lambda_{\text{BNV}}$ with $\chi_{_{uu}}$ mass. We fix $\lambda_{LLLQ}$ = 1, set $m_{\chis{LL}} = m_{\chis{LQ}} = m$ = 1.5 TeV, $| y_{uc} | $ = 0.25, $\vert y_{ut} \vert $ = $\vert y_{ct} \vert $ = 0.12 and show the resulting constraints for $\sqrt{s}=\sqrt{4.5}m$ TeV (light green) and $\sqrt{s}=\sqrt{15}m$ (light blue). We always evaluate the perturbative unitarity constraints at  $\sqrt{s}=\sqrt{4.5}m$ but show the constraints for $\sqrt{s}=\sqrt{4.5}m$ to show the effect of increasing $s$. The ``funnel" shape of these contours is due to resonance in the $s$ channel diagram, as explained in more detail in the text. We also show contours of values of $\Lambda_{\text{BNV}}$ for which the branching fraction of the $\chis{uu}^\dagger \rightarrow \chis{LQ}^\dagger \ \chis{LL}$ mode is 0.1 (blue) and 0.5 (dark red) respectively to illustrate that for much of the mass range for the diquark considered in this work, perturbative unitarity allows a substantial fraction of the diquark decays to be to the BNV mode. }
\label{fig:PerturbativeUnitarityLimits}
\end{figure}

To determine the perturbative unitarity constraints on $ \Lambda_{\text{BNV}}$ and $\lambda_{LQLL}$, we apply the general formulae provided in \cite{Goodsell:2018tti}, but with slight modifications to deal with poles in the propagator of the diquark. The amplitude can thus be written as
\begin{widetext}
\begin{equation}\label{eq:partialwaveunitarityamplitude}
    \mathcal{M}(\chi_{_{LL}}\chi_{_{LQ}}\rightarrow\chi_{_{LL}}\chi_{_{LQ}}) = \lambda_{_{LLLQ}} + \frac{\Lambda^{2}_{\text{BNV}}}{s-m^{2}_{\chis{uu}}+im_{\chis{uu}}\Gamma_{uu}} + \frac{\Lambda^{2}_{\text{BNV}}}{t-m^{2}_{\chis{uu}}+im_{\chis{uu}}\Gamma_{uu}} + \frac{\Lambda^{2}_{\text{BNV}}}{u-m^{2}_{\chis{uu}}+im_{\chis{uu}}\Gamma_{uu}},
\end{equation}    
\end{widetext}
where $\Gamma_{uu}$ is the decay width of $\chi_{_{uu}}$ and $s,t,$ and $u$ are the usual Mandelstam variables. The partial wave unitarity constraint is defined by the inequality
\begin{equation}
    -i\left(a_{j} - a^{\dagger}_{j}\right) \leq a_{j}a^{\dagger}_{j},
\end{equation}
where $a_{j}$ is the $j^{\text{th}}$ coefficient of the partial wave decomposition of the amplitude. Since this must be satisfied term by term, the authors of \cite{Goodsell:2018tti} consider only the zeroth order one and find,
\begin{equation}\label{eq:perturbativeunitaritycondition}
    \text{Re}(a_{0}) \leq \frac{1}{2}.
\end{equation}
We apply the same conditions and, in the $s \rightarrow \infty$ limit, find
\begin{equation}
\lambda_{LQLL} > -8\pi,    
\end{equation}
a constraint independent of $\Lambda_{\text{BNV}}$ that is easily satisfied by the previously imposed positivity criterion to avoid spontaneous breaking of electromagnetic gauge invariance. As this asymptotic limit implies, the perturbative unitarity constraints on $ \Lambda_{\text{BNV}}$ weakens with increasing $s$. However, it must hold for \emph{all} $s$ above the kinematic threshold for the scattering process in question. To derive the limits for for different values of $s$, we apply the formulae presented in \cite{Goodsell:2018tti}, but with the previously mentioned modification to deal with $s$ channel poles that effectively amounts to the replacement 
\begin{equation*}
m^{2}_{\chis{uu}} \rightarrow m^{2}_{\chis{uu}} + im_{\chis{uu}}\Gamma_{uu}.    
\end{equation*}
We fix the masses $m_{\chis{LL}}= m_{\chis{LQ}} = m = 1.5$ TeV, consistent with existing constraints surveyed in Section~\eqref{sec:ExistingPheno}. For the center of mass energy, we use $\sqrt{s} = \sqrt{4.5}\times m$, just slightly above the kinematic threshold. The results are shown in Figure~\eqref{fig:PerturbativeUnitarityLimits}, where the shaded green region violates perturbative unitarity. For illustrative purposes, we also show the effect of increasing $s$ by plotting the exclusion limit at $\sqrt{s} = \sqrt{15}\times m$ in blue. The shape of the countours can be understood by noting that the partial wave coefficients $a_{j}$ are proportional to the amplitude for the scattering process. As shown in equation~\eqref{eq:partialwaveunitarityamplitude}, this increases due to the $s$-channel resonance when $s$ approaches $m^{2}_{\chis{uu}}$, leading to the funnel shaped contours. The increasing amplitude also means that the constraint~\eqref{eq:perturbativeunitaritycondition} can be satisfied for smaller values of $\Lambda_{\rm BNV}$. Since the green and the blue regions correspond to different values of $s$, the ``funnelling" happens at different $m_{\chi_{_{uu}}}$, with that of the blue contour being higher due to the higher $s$. On the same figure, we also show values of the trilinear coupling $\Lambda_{\text{BNV}}$ for which the BNV branching ratio of $\chi_{_{uu}}$,  $\text{Br}_{\text{BNV}}$, is 10$\%$ (dark blue) and 50$\%$ (dark red) of the total its total decay width for a choice of Yukawa couplings $| y_{uc} | $ = 0.25 and  $| y_{ut} | $ = $| y_{ct} | $ = 0.12, values that are consistent with the limits presented in~\eqref{sec:ExistingPheno}. Importantly, as Figure~\eqref{fig:PerturbativeUnitarityLimits} illustrates, for much of the parameter space that remains unconstrained by perturbative unitarity, one can get relatively large fractions of the diquark decays to be to the BNV mode, which is advantageous for the for searches of this model at colliders.   

\newpage
\section{Calculation of the Proton Decay Rate}\label{sec:PDDetailedCalculations}
Averaged over the spin states of the proton, the differential decay rate for the process $P(p)\rightarrow \overline{\nu}_{l}(k_{1}) \overline{\nu}_{m}(k_{2}) l^{+}_{n} (k_{3})$ ($l,m,n$ are lepton flavour indices and $m \neq n$) is 
\begin{equation}
    d\Gamma = \frac{1}{\left(2\pi\right)^{3}}\frac{1}{32m^{3}_{P}} \vert \overline{\mathcal{M}} \vert^{2}_{p \rightarrow \overline{l}\overline{\nu}\overline{\nu}} \ dm^{2}_{12} \ dm^{2}_{23}.
\end{equation}
Here,
\begin{equation}
\begin{split}
    m^{2}_{12} & = \left(k_{1} + k_{2}\right)^{2} \\ 
    m^{2}_{23} & = \left(k_{2} + k_{3}\right)^{2},
\end{split}
\end{equation}
are kinematic invariants and $\vert \overline{\mathcal{M}}^{2} \vert_{p \rightarrow \overline{l}\overline{\nu\nu}}$ is the spin averaged square of the amplitude for the proton decay process. We will assume throughout that $m_{\nu}$ = 0 and average over outgoing neutrino flavours. \\

For a specific flavour combination of virtual fermions, there are two loop diagrams for proton decay, as shown in Fig.~\eqref{fig:ProtonDecayUVComplete}. If the outgoing antineutrino flavours $ l= m$, each of these diagrams will have a companion with $k_{1} \leftrightarrow k_{2}$. Schematically, we can split the amplitude as 
\begin{equation}
    \mathcal{M}_{p \rightarrow \overline{l}\overline{\nu}\overline{\nu}} = \mathcal{M}^{q}_{p \rightarrow \overline{l}\overline{\nu}\overline{\nu}} + \mathcal{M}^{l}_{p \rightarrow \overline{l}\overline{\nu}\overline{\nu}}, 
\end{equation}
where $\mathcal{M}^{q}_{p \rightarrow \overline{l}\overline{\nu}\overline{\nu}}$ corresponds to the diagram shown on the upper panel of \cref{fig:ProtonDecayUVComplete} and $\mathcal{M}^{l}_{p \rightarrow \overline{l}\overline{\nu}\overline{\nu}}$ corresponds to the lower panel. The loop integral, after simplifying the numerator and using Feynman parameterization, is given by
\begin{widetext}
\begin{align}
   I = -2\times 4! \times m_{u_{k}}P_{L} \gamma^{\nu} \int_{0}^{1} dx_{1} \int_{0}^{1-x_{1}} dx_{2} \int_{0}^{1-\sum\limits_{j=1}^{2}x_j} dx_{3} \int_{0}^{1-\sum\limits_{j=1}^{3}x_j} dx_{4}\nonumber\int \frac{d^{D}q}{(2\pi)^{D}} \frac{q_{\nu}-x_{1}p_{2\nu}-x_{2}(p_{1\nu}+p_{2\nu}) - x_{3}p_{\nu} - x_{4}k_{1\nu}}{\left(q^{2}-\Delta_q^{2}\right)^{5}}, 
\end{align}    
with 
\begin{equation}
\begin{split}
   \Delta^{2}_{q} & = \left(x_{1}p_{2}+x_{2}(p_{1}+p_{2}) + x_{3}p+x_{4}k_{1}\right)^{2}  
    +  x_{1}(m_{W}-p^{2}_{2}) + x_{2}(m^{2}_{u_{k}}-(p_{1}+p_{2})^{2})+ x_{3}(m^{2}_{\chis{uu}}-p^{2}) + x_{4}(m^{2}_{\chis{LQ}}-k_{1}^{2}) \\
    & - (x_{1}+x_{2}+x_{3}+x_{4})m^{2}_{d_{j}}, 
\end{split}
\end{equation}
\end{widetext}
where $p = p_{1} + p_{2} + p_{3}$ is the total proton four momentum. The first term integrates to zero by symmetry while the remaining ones are given by 
\begin{widetext}
\begin{equation}
\begin{split}
   I = \frac{-2m_{u_{k}}P_{L}}{16\pi^{2}} \int_{0}^{1} dx_{1} \int_{0}^{1-x_{1}} dx_{2} \int_{0}^{1-x_{1}-x_{2}} dx_{3} \int_{0}^{1-x_{1}-x_{2}-x_{3}} dx_{4} \ \frac{x_{1}\slashed{p}_{2}+x_{2}(\slashed{p}_{1}+\slashed{p}_{2}) + x_{3}\slashed{p} + x_{4}\slashed{k}_{1}}{\left(\Delta^{2}_{q}\right)^{3}}, 
\end{split}
\end{equation}    
\end{widetext}
which can be evaluated numerically.  Omitting the coupling constants and the leptonic contribution from the dilepton singlet temporarily to avoid notational cluttering, we factorize the remaining terms in the amplitude into the leptonic and hadronic contributions as,  
\begin{widetext}
\begin{equation}\label{eq:protondecayspinors}
    \left( C_{1}\slashed{p_{2}} + C_{2}(\slashed{p_{1}}+\slashed{p_{2}}) + C_{3}\slashed{p} + C_{4}(\slashed{p}-\slashed{k}_{2}-\slashed{k}_{3}) \right) \times \left< 0 \vert \left(\overline{d}^{c}(p_{1})P_{L}u(p_{3})\right) \overline{u}^{c}_{L}(p_{2}) \vert p \right> P_{L} v_{l}(k_{1}) ,
\end{equation}    
\end{widetext}
where $C_{1,2,3,4}$ are the coefficients calculated from the loop integral, and we have written $\slashed{k}_{1}$ as $\slashed{p} - \slashed{k}_{2} -\slashed{k}_{3}$ to extract every possible action of the quark ``slashed momentum'' operators on their spinor fields. This algebraic manipulation is important in order to use chirality to carefully identify non-vanishing and vanishing contributions to the amplitude.\\

To proceed, some manipulations of the hadronic portion of the proton decay matrix element are necessary: we need to relate it to the proton to vacuum matrix element in~\eqref{eq:protondecayspinors}. This is the three quark BNV operator 
that the authors write in shorthand form (see their equation (10) as
\begin{subequations}
 \begin{equation}\label{eq:AokiPDMatrixElement1}
    \left< 0 \ \vert (ud)_{R}u_{L} \vert p \right> = \alpha P_{R} u_{P}(p) 
\end{equation}
\begin{equation}\label{eq:AokiPDMatrixElement2}
 \left< 0 \ \vert (ud)_{L}u_{L} \vert p \right> = \beta P_{L} u_{P}(p),
\end{equation}
\end{subequations}
with the results of the lattice computations in \cite{Aoki:2017puj} giving $\alpha$ = $-\beta = -0.014$ GeV$^{3}$ (evaluated at a scale $\mu$ = 2 GeV).
Noting that
\begin{equation*}
    \begin{split}
      \left( \overline{d}^{c}P_{L}u\right) & = -\left( \overline{d}^{c}P_{L}u\right)^{T} = u^{T}C P_{L}d 
    \end{split}
\end{equation*}
and using 
\begin{equation*}
\overline{u}^{c} = - u^{T}C^{-1},    
\end{equation*}
we apply transposition to the vacuum proton decay matrix element and perform some further algebra to find
\begin{equation}
    \langle \  0 \ \vert  \left(\overline{d}^{c}P_{L}u\right)\overline{u}^{c}_{L} \vert \ p \ \rangle = \beta P_{L} \overline{v}_{P}(p).
\end{equation}
This is consistent with the form of the matrix elements used in \cite{Hou:2005iu} (see, for example, just below equation (19) for neutron to vacuum matrix element). \\

Applying the on shell condition for the valence quarks, we see, for example, that $\overline{u}^{c}_{L}\slashed{p}_{2} = m^{*}_{u} \ \overline{u}^{c}_{R}$\footnote{We use a * to distinguish the fact that we are using the constituent quark mass, appropriate for quarks inside a proton. See the main text for details.} and simplifying using the matrix elements~\eqref{eq:AokiPDMatrixElement2}
\begin{widetext}
\begin{equation*}
\begin{split}
   v_{l}(k_{1}) P_{L} \times \left( C_{1}\slashed{p_{2}} \right)\times \left< \ 0 \ \vert \left(\overline{d}^{c}(p_{1})_{L}u(p_{3})\right)\overline{u}^{c}_{L}(p_{2}) \vert p \right> & = v_{l}(k_{1}) C_{1}m_{u}P_{L}  \left< 0 \vert \left(\overline{d}^{c}(p_{1})P_{L}u(p_{3})\right)\overline{u}^{c}_{R}(p_{2}) \vert p \right> \\ 
   & = \alpha \times C_{1} \times m_{u} \overline{v}_{P}(p)P_{L}v_{l}(k_{1})  
\end{split}
\end{equation*}    
\end{widetext}
This term survives. However, if we now apply $ \overline{d}^{c}_{L}\slashed{p}_{1} = m^{*}_{d} \overline{d}^{c}_{R}$ and $\slashed p_{3} u_{L}(p_{3}) = m^{*}_{u} u_{R}(p_{3})$, an analogous exercise shows that these terms vanish. Therefore, the only contribution from the matrix element is in the form of $\left(C_{1} + C_{2} + C_{3} + C_{4}\right)\slashed{p}_{2} \equiv C^{q}_{\text{loop}} \slashed{p}_{2} $. With these, we can write the final amplitude as
\begin{widetext}
\begin{equation}\label{eq:ampprotondecayquark}
        \mathcal{M}^{q}_{p \rightarrow \overline{l\nu\nu}} = 32\alpha_{\text{QCD}} \frac{e^{2}}{s^{2}_{w}} \frac{\lambda^{*}\left<\Lambda_{B}\right>}{m^{2}_{LL}} \times \overline{v}_{l_{2}}(k_{2}) P_{L} v_{l_{3}}(k_{3}) \overline{v}_{P}(p) P_{L} v_{l_{1}} (k_{1}) \times \sum_{j} \sum_{k} m_{u_{k}}m^{*}_{u} \times y^{l_{1}j}_{LQ}y^{ku}_{uu}y^{l_{2}l_{3}}_{LL} \times V^{*uk}V^{jd} \times C^{q}_{\text{loop}}, 
\end{equation}
\end{widetext}
where $V$ is the CKM matrix and the constant $\alpha_{\text{QCD}}= 0.014$ GeV$^{3}$ is taken from the lattice computations reported in \cite{Aoki:2017puj}. The $1/m^{2}_{\chis{LL}}$ term comes from approximating the $\chis{LL}$ propagator as $1/m^{2}_{\chis{LL}}$ and neglecting the $\mathcal{O}(m^{2}_{P})$ momentum term, with 
\begin{widetext}
\begin{equation}\label{eq:Cqloop}
\begin{split}
   C^{q}_{\text{loop}} = \frac{N_{c}}{16\pi^{2}} \int_{0}^{1} dx_{1} \int_{0}^{1-x_{1}} dx_{2} \int_{0}^{1-x_{1}-x_{2}} dx_{3} \int_{0}^{1-x_{1}-x_{2}-x_{3}} dx_{4} \ \frac{x_{1}+x_{2} + x_{3} + x_{4}}{\left(\Delta^{2}_{q}\right)^{3}}, 
\end{split}
\end{equation}    
\end{widetext} 
and $N_{c} = 3$ is the number of colours for the intermediate up type quark. An analogous exercise for the second diagram gives
\begin{widetext}
\begin{equation}\label{eq:ampprotondecaylepton} 
        \mathcal{M}^{l}_{p \rightarrow \overline{l\nu\nu}} = -32\alpha_{\text{QCD}} \frac{e^{2}}{s^{2}_{w}} \frac{\lambda^{*}\left<\Lambda_{B}\right>}{m^{2}_{LL}} \times \overline{v}_{l_{2}}(k_{2}) P_{L} v_{l_{3}}(k_{3}) \overline{v}_{P}(p) P_{L} v_{l_{1}} (k_{1}) \times \sum_{j} \sum_{k} m_{u_{k}}m^{*}_{u} \times y^{l_{1}j}_{LQ}y^{ku}_{uu}y^{l_{2}l_{3}}_{LL} \times V^{*uk}U^{jl_{1}} \times C^{l}_{\text{loop}}, 
\end{equation}   
\end{widetext}
where we note that due to the relative sign difference of the lepto-quark couplings between up-type quarks to charged leptons and that between down-type quarks to neutrinos there is a sign difference at the level of the \emph{amplitude}, leading to destructive interference between the two diagrams. $U$ denotes the PMNS matrix and 
\begin{widetext}
\begin{equation}\label{eq:Clloop}
\begin{split}
   C^{l}_{\text{loop}} = \frac{N_{c}}{16\pi^{2}} \int_{0}^{1} dx_{1} \int_{0}^{1-x_{1}} dx_{2} \int_{0}^{1-x_{1}-x_{2}} dx_{3} \int_{0}^{1-x_{1}-x_{2}-x_{3}} dx_{4} \ \frac{x_{1} + x_{2} + x_{3} + x_{4}}{\left(\Delta^{2}_{l}\right)^{3}}, 
\end{split}
\end{equation}    
with
\begin{equation}
\begin{split}
   \Delta^{2}_{l} & = \left(x_{1}p_{2}+x_{2}(p_{2}-k_{2}-k_{3}) + x_{3}(p_{1}+p_{2}-k_{2}-k_{3})+x_{4}k_{1}\right)^{2} + \\ 
    & +  x_{1}(m^{2}_{\chis{LQ}}-p^{2}_{2}) + x_{2}(m^{2}_{\chis{uu}}-(p_{2}-k_{2}-k_{3})^{2}) + x_{3}(m^{2}_{u_{k}}-(p_{3}-k_{1})^{2}) + x_{4}(m^{2}_{W}-k_{1}^{2}) - (x_{1}+x_{2}+x_{3}+x_{4}-1)m^{2}_{l}. 
\end{split}
\end{equation}
\end{widetext}

\bibliography{apssamp}

%merlin.mbs apsrev4-1.bst 2010-07-25 4.21a (PWD, AO, DPC) hacked
%Control: key (0)
%Control: author (8) initials jnrlst
%Control: editor formatted (1) identically to author
%Control: production of article title (-1) disabled
%Control: page (0) single
%Control: year (1) truncated
%Control: production of eprint (0) enabled
\begin{thebibliography}{70}%
\makeatletter
\providecommand \@ifxundefined [1]{%
 \@ifx{#1\undefined}
}%
\providecommand \@ifnum [1]{%
 \ifnum #1\expandafter \@firstoftwo
 \else \expandafter \@secondoftwo
 \fi
}%
\providecommand \@ifx [1]{%
 \ifx #1\expandafter \@firstoftwo
 \else \expandafter \@secondoftwo
 \fi
}%
\providecommand \natexlab [1]{#1}%
\providecommand \enquote  [1]{``#1''}%
\providecommand \bibnamefont  [1]{#1}%
\providecommand \bibfnamefont [1]{#1}%
\providecommand \citenamefont [1]{#1}%
\providecommand \href@noop [0]{\@secondoftwo}%
\providecommand \href [0]{\begingroup \@sanitize@url \@href}%
\providecommand \@href[1]{\@@startlink{#1}\@@href}%
\providecommand \@@href[1]{\endgroup#1\@@endlink}%
\providecommand \@sanitize@url [0]{\catcode `\\12\catcode `\$12\catcode `\&12\catcode `\#12\catcode `\^12\catcode `\_12\catcode `\%12\relax}%
\providecommand \@@startlink[1]{}%
\providecommand \@@endlink[0]{}%
\providecommand \url  [0]{\begingroup\@sanitize@url \@url }%
\providecommand \@url [1]{\endgroup\@href {#1}{\urlprefix }}%
\providecommand \urlprefix  [0]{URL }%
\providecommand \Eprint [0]{\href }%
\providecommand \doibase [0]{http://dx.doi.org/}%
\providecommand \selectlanguage [0]{\@gobble}%
\providecommand \bibinfo  [0]{\@secondoftwo}%
\providecommand \bibfield  [0]{\@secondoftwo}%
\providecommand \translation [1]{[#1]}%
\providecommand \BibitemOpen [0]{}%
\providecommand \bibitemStop [0]{}%
\providecommand \bibitemNoStop [0]{.\EOS\space}%
\providecommand \EOS [0]{\spacefactor3000\relax}%
\providecommand \BibitemShut  [1]{\csname bibitem#1\endcsname}%
\let\auto@bib@innerbib\@empty
%</preamble>
\bibitem [{\citenamefont {Abbott}\ and\ \citenamefont {Wise}(1989)}]{Abbott:1989jw}%
  \BibitemOpen
  \bibfield  {author} {\bibinfo {author} {\bibfnamefont {L.~F.}\ \bibnamefont {Abbott}}\ and\ \bibinfo {author} {\bibfnamefont {M.~B.}\ \bibnamefont {Wise}},\ }\href {\doibase 10.1016/0550-3213(89)90503-8} {\bibfield  {journal} {\bibinfo  {journal} {Nucl. Phys. B}\ }\textbf {\bibinfo {volume} {325}},\ \bibinfo {pages} {687} (\bibinfo {year} {1989})}\BibitemShut {NoStop}%
\bibitem [{\citenamefont {Kallosh}\ \emph {et~al.}(1995)\citenamefont {Kallosh}, \citenamefont {Linde}, \citenamefont {Linde},\ and\ \citenamefont {Susskind}}]{Kallosh:1995hi}%
  \BibitemOpen
  \bibfield  {author} {\bibinfo {author} {\bibfnamefont {R.}~\bibnamefont {Kallosh}}, \bibinfo {author} {\bibfnamefont {A.~D.}\ \bibnamefont {Linde}}, \bibinfo {author} {\bibfnamefont {D.~A.}\ \bibnamefont {Linde}}, \ and\ \bibinfo {author} {\bibfnamefont {L.}~\bibnamefont {Susskind}},\ }\href {\doibase 10.1103/PhysRevD.52.912} {\bibfield  {journal} {\bibinfo  {journal} {Phys. Rev. D}\ }\textbf {\bibinfo {volume} {52}},\ \bibinfo {pages} {912} (\bibinfo {year} {1995})},\ \Eprint {http://arxiv.org/abs/hep-th/9502069} {arXiv:hep-th/9502069} \BibitemShut {NoStop}%
\bibitem [{\citenamefont {Lee}(1988)}]{Lee:1988ge}%
  \BibitemOpen
  \bibfield  {author} {\bibinfo {author} {\bibfnamefont {K.-M.}\ \bibnamefont {Lee}},\ }\href {\doibase 10.1103/PhysRevLett.61.263} {\bibfield  {journal} {\bibinfo  {journal} {Phys. Rev. Lett.}\ }\textbf {\bibinfo {volume} {61}},\ \bibinfo {pages} {263} (\bibinfo {year} {1988})}\BibitemShut {NoStop}%
\bibitem [{\citenamefont {Sakharov}(1967)}]{Sakharov:1967dj}%
  \BibitemOpen
  \bibfield  {author} {\bibinfo {author} {\bibfnamefont {A.~D.}\ \bibnamefont {Sakharov}},\ }\href {\doibase 10.1070/PU1991v034n05ABEH002497} {\bibfield  {journal} {\bibinfo  {journal} {Pisma Zh. Eksp. Teor. Fiz.}\ }\textbf {\bibinfo {volume} {5}},\ \bibinfo {pages} {32} (\bibinfo {year} {1967})}\BibitemShut {NoStop}%
\bibitem [{\citenamefont {Davis}\ \emph {et~al.}(1968)\citenamefont {Davis}, \citenamefont {Harmer},\ and\ \citenamefont {Hoffman}}]{Davis:1968cp}%
  \BibitemOpen
  \bibfield  {author} {\bibinfo {author} {\bibfnamefont {R.}~\bibnamefont {Davis}, \bibfnamefont {Jr.}}, \bibinfo {author} {\bibfnamefont {D.~S.}\ \bibnamefont {Harmer}}, \ and\ \bibinfo {author} {\bibfnamefont {K.~C.}\ \bibnamefont {Hoffman}},\ }\href {\doibase 10.1103/PhysRevLett.20.1205} {\bibfield  {journal} {\bibinfo  {journal} {Phys. Rev. Lett.}\ }\textbf {\bibinfo {volume} {20}},\ \bibinfo {pages} {1205} (\bibinfo {year} {1968})}\BibitemShut {NoStop}%
\bibitem [{\citenamefont {Fukuda}\ \emph {et~al.}(1998)\citenamefont {Fukuda} \emph {et~al.}}]{Super-Kamiokande:1998kpq}%
  \BibitemOpen
  \bibfield  {author} {\bibinfo {author} {\bibfnamefont {Y.}~\bibnamefont {Fukuda}} \emph {et~al.} (\bibinfo {collaboration} {Super-Kamiokande}),\ }\href {\doibase 10.1103/PhysRevLett.81.1562} {\bibfield  {journal} {\bibinfo  {journal} {Phys. Rev. Lett.}\ }\textbf {\bibinfo {volume} {81}},\ \bibinfo {pages} {1562} (\bibinfo {year} {1998})},\ \Eprint {http://arxiv.org/abs/hep-ex/9807003} {arXiv:hep-ex/9807003} \BibitemShut {NoStop}%
\bibitem [{\citenamefont {Ahmad}\ \emph {et~al.}(2001)\citenamefont {Ahmad} \emph {et~al.}}]{SNO:2001kpb}%
  \BibitemOpen
  \bibfield  {author} {\bibinfo {author} {\bibfnamefont {Q.~R.}\ \bibnamefont {Ahmad}} \emph {et~al.} (\bibinfo {collaboration} {SNO}),\ }\href {\doibase 10.1103/PhysRevLett.87.071301} {\bibfield  {journal} {\bibinfo  {journal} {Phys. Rev. Lett.}\ }\textbf {\bibinfo {volume} {87}},\ \bibinfo {pages} {071301} (\bibinfo {year} {2001})},\ \Eprint {http://arxiv.org/abs/nucl-ex/0106015} {arXiv:nucl-ex/0106015} \BibitemShut {NoStop}%
\bibitem [{\citenamefont {Takenaka}\ \emph {et~al.}(2020)\citenamefont {Takenaka} \emph {et~al.}}]{Super-Kamiokande:2020wjk}%
  \BibitemOpen
  \bibfield  {author} {\bibinfo {author} {\bibfnamefont {A.}~\bibnamefont {Takenaka}} \emph {et~al.} (\bibinfo {collaboration} {Super-Kamiokande}),\ }\href {\doibase 10.1103/PhysRevD.102.112011} {\bibfield  {journal} {\bibinfo  {journal} {Phys. Rev. D}\ }\textbf {\bibinfo {volume} {102}},\ \bibinfo {pages} {112011} (\bibinfo {year} {2020})},\ \Eprint {http://arxiv.org/abs/2010.16098} {arXiv:2010.16098 [hep-ex]} \BibitemShut {NoStop}%
\bibitem [{\citenamefont {McGrew}\ \emph {et~al.}(1999)\citenamefont {McGrew} \emph {et~al.}}]{McGrew:1999nd}%
  \BibitemOpen
  \bibfield  {author} {\bibinfo {author} {\bibfnamefont {C.}~\bibnamefont {McGrew}} \emph {et~al.},\ }\href {\doibase 10.1103/PhysRevD.59.052004} {\bibfield  {journal} {\bibinfo  {journal} {Phys. Rev. D}\ }\textbf {\bibinfo {volume} {59}},\ \bibinfo {pages} {052004} (\bibinfo {year} {1999})}\BibitemShut {NoStop}%
\bibitem [{\citenamefont {Hirata}\ \emph {et~al.}(1989)\citenamefont {Hirata} \emph {et~al.}}]{Kamiokande-II:1989avz}%
  \BibitemOpen
  \bibfield  {author} {\bibinfo {author} {\bibfnamefont {K.~S.}\ \bibnamefont {Hirata}} \emph {et~al.} (\bibinfo {collaboration} {Kamiokande-II}),\ }\href {\doibase 10.1016/0370-2693(89)90058-0} {\bibfield  {journal} {\bibinfo  {journal} {Phys. Lett. B}\ }\textbf {\bibinfo {volume} {220}},\ \bibinfo {pages} {308} (\bibinfo {year} {1989})}\BibitemShut {NoStop}%
\bibitem [{\citenamefont {Abe}\ \emph {et~al.}(2014)\citenamefont {Abe} \emph {et~al.}}]{Super-Kamiokande:2014otb}%
  \BibitemOpen
  \bibfield  {author} {\bibinfo {author} {\bibfnamefont {K.}~\bibnamefont {Abe}} \emph {et~al.} (\bibinfo {collaboration} {Super-Kamiokande}),\ }\href {\doibase 10.1103/PhysRevD.90.072005} {\bibfield  {journal} {\bibinfo  {journal} {Phys. Rev. D}\ }\textbf {\bibinfo {volume} {90}},\ \bibinfo {pages} {072005} (\bibinfo {year} {2014})},\ \Eprint {http://arxiv.org/abs/1408.1195} {arXiv:1408.1195 [hep-ex]} \BibitemShut {NoStop}%
\bibitem [{\citenamefont {Kobayashi}\ \emph {et~al.}(2005)\citenamefont {Kobayashi} \emph {et~al.}}]{Super-Kamiokande:2005lev}%
  \BibitemOpen
  \bibfield  {author} {\bibinfo {author} {\bibfnamefont {K.}~\bibnamefont {Kobayashi}} \emph {et~al.} (\bibinfo {collaboration} {Super-Kamiokande}),\ }\href {\doibase 10.1103/PhysRevD.72.052007} {\bibfield  {journal} {\bibinfo  {journal} {Phys. Rev. D}\ }\textbf {\bibinfo {volume} {72}},\ \bibinfo {pages} {052007} (\bibinfo {year} {2005})},\ \Eprint {http://arxiv.org/abs/hep-ex/0502026} {arXiv:hep-ex/0502026} \BibitemShut {NoStop}%
\bibitem [{\citenamefont {Heeck}\ and\ \citenamefont {Takhistov}(2020)}]{Heeck:2019kgr}%
  \BibitemOpen
  \bibfield  {author} {\bibinfo {author} {\bibfnamefont {J.}~\bibnamefont {Heeck}}\ and\ \bibinfo {author} {\bibfnamefont {V.}~\bibnamefont {Takhistov}},\ }\href {\doibase 10.1103/PhysRevD.101.015005} {\bibfield  {journal} {\bibinfo  {journal} {Phys. Rev. D}\ }\textbf {\bibinfo {volume} {101}},\ \bibinfo {pages} {015005} (\bibinfo {year} {2020})},\ \Eprint {http://arxiv.org/abs/1910.07647} {arXiv:1910.07647 [hep-ph]} \BibitemShut {NoStop}%
\bibitem [{\citenamefont {Georgi}\ and\ \citenamefont {Glashow}(1974)}]{Georgi:1974sy}%
  \BibitemOpen
  \bibfield  {author} {\bibinfo {author} {\bibfnamefont {H.}~\bibnamefont {Georgi}}\ and\ \bibinfo {author} {\bibfnamefont {S.~L.}\ \bibnamefont {Glashow}},\ }\href {\doibase 10.1103/PhysRevLett.32.438} {\bibfield  {journal} {\bibinfo  {journal} {Phys. Rev. Lett.}\ }\textbf {\bibinfo {volume} {32}},\ \bibinfo {pages} {438} (\bibinfo {year} {1974})}\BibitemShut {NoStop}%
\bibitem [{\citenamefont {Pati}\ and\ \citenamefont {Salam}(1974)}]{Pati:1974yy}%
  \BibitemOpen
  \bibfield  {author} {\bibinfo {author} {\bibfnamefont {J.~C.}\ \bibnamefont {Pati}}\ and\ \bibinfo {author} {\bibfnamefont {A.}~\bibnamefont {Salam}},\ }\href {\doibase 10.1103/PhysRevD.10.275} {\bibfield  {journal} {\bibinfo  {journal} {Phys. Rev. D}\ }\textbf {\bibinfo {volume} {10}},\ \bibinfo {pages} {275} (\bibinfo {year} {1974})},\ \bibinfo {note} {[Erratum: Phys.Rev.D 11, 703--703 (1975)]}\BibitemShut {NoStop}%
\bibitem [{\citenamefont {Fritzsch}\ and\ \citenamefont {Minkowski}(1975)}]{Fritzsch:1974nn}%
  \BibitemOpen
  \bibfield  {author} {\bibinfo {author} {\bibfnamefont {H.}~\bibnamefont {Fritzsch}}\ and\ \bibinfo {author} {\bibfnamefont {P.}~\bibnamefont {Minkowski}},\ }\href {\doibase 10.1016/0003-4916(75)90211-0} {\bibfield  {journal} {\bibinfo  {journal} {Annals Phys.}\ }\textbf {\bibinfo {volume} {93}},\ \bibinfo {pages} {193} (\bibinfo {year} {1975})}\BibitemShut {NoStop}%
\bibitem [{\citenamefont {Weinberg}(1980)}]{Weinberg:1980bf}%
  \BibitemOpen
  \bibfield  {author} {\bibinfo {author} {\bibfnamefont {S.}~\bibnamefont {Weinberg}},\ }\href {\doibase 10.1103/PhysRevD.22.1694} {\bibfield  {journal} {\bibinfo  {journal} {Phys. Rev. D}\ }\textbf {\bibinfo {volume} {22}},\ \bibinfo {pages} {1694} (\bibinfo {year} {1980})}\BibitemShut {NoStop}%
\bibitem [{\citenamefont {Wilczek}\ and\ \citenamefont {Zee}(1979)}]{Wilczek:1979hc}%
  \BibitemOpen
  \bibfield  {author} {\bibinfo {author} {\bibfnamefont {F.}~\bibnamefont {Wilczek}}\ and\ \bibinfo {author} {\bibfnamefont {A.}~\bibnamefont {Zee}},\ }\href {\doibase 10.1103/PhysRevLett.43.1571} {\bibfield  {journal} {\bibinfo  {journal} {Phys. Rev. Lett.}\ }\textbf {\bibinfo {volume} {43}},\ \bibinfo {pages} {1571} (\bibinfo {year} {1979})}\BibitemShut {NoStop}%
\bibitem [{\citenamefont {Abe}\ \emph {et~al.}(2018)\citenamefont {Abe} \emph {et~al.}}]{Hyper-Kamiokande:2018ofw}%
  \BibitemOpen
  \bibfield  {author} {\bibinfo {author} {\bibfnamefont {K.}~\bibnamefont {Abe}} \emph {et~al.} (\bibinfo {collaboration} {Hyper-Kamiokande}),\ }\href@noop {} {\  (\bibinfo {year} {2018})},\ \Eprint {http://arxiv.org/abs/1805.04163} {arXiv:1805.04163 [physics.ins-det]} \BibitemShut {NoStop}%
\bibitem [{\citenamefont {Takhistov}\ \emph {et~al.}(2014)\citenamefont {Takhistov} \emph {et~al.}}]{Super-Kamiokande:2014pqx}%
  \BibitemOpen
  \bibfield  {author} {\bibinfo {author} {\bibfnamefont {V.}~\bibnamefont {Takhistov}} \emph {et~al.} (\bibinfo {collaboration} {Super-Kamiokande}),\ }\href {\doibase 10.1103/PhysRevLett.113.101801} {\bibfield  {journal} {\bibinfo  {journal} {Phys. Rev. Lett.}\ }\textbf {\bibinfo {volume} {113}},\ \bibinfo {pages} {101801} (\bibinfo {year} {2014})},\ \Eprint {http://arxiv.org/abs/1409.1947} {arXiv:1409.1947 [hep-ex]} \BibitemShut {NoStop}%
\bibitem [{\citenamefont {Fonseca}\ \emph {et~al.}(2018)\citenamefont {Fonseca}, \citenamefont {Hirsch},\ and\ \citenamefont {Srivastava}}]{Fonseca:2018ehk}%
  \BibitemOpen
  \bibfield  {author} {\bibinfo {author} {\bibfnamefont {R.~M.}\ \bibnamefont {Fonseca}}, \bibinfo {author} {\bibfnamefont {M.}~\bibnamefont {Hirsch}}, \ and\ \bibinfo {author} {\bibfnamefont {R.}~\bibnamefont {Srivastava}},\ }\href {\doibase 10.1103/PhysRevD.97.075026} {\bibfield  {journal} {\bibinfo  {journal} {Phys. Rev. D}\ }\textbf {\bibinfo {volume} {97}},\ \bibinfo {pages} {075026} (\bibinfo {year} {2018})},\ \Eprint {http://arxiv.org/abs/1802.04814} {arXiv:1802.04814 [hep-ph]} \BibitemShut {NoStop}%
\bibitem [{\citenamefont {Babu}\ \emph {et~al.}(2006)\citenamefont {Babu}, \citenamefont {Mohapatra},\ and\ \citenamefont {Nasri}}]{Babu:2006xc}%
  \BibitemOpen
  \bibfield  {author} {\bibinfo {author} {\bibfnamefont {K.~S.}\ \bibnamefont {Babu}}, \bibinfo {author} {\bibfnamefont {R.~N.}\ \bibnamefont {Mohapatra}}, \ and\ \bibinfo {author} {\bibfnamefont {S.}~\bibnamefont {Nasri}},\ }\href {\doibase 10.1103/PhysRevLett.97.131301} {\bibfield  {journal} {\bibinfo  {journal} {Phys. Rev. Lett.}\ }\textbf {\bibinfo {volume} {97}},\ \bibinfo {pages} {131301} (\bibinfo {year} {2006})},\ \Eprint {http://arxiv.org/abs/hep-ph/0606144} {arXiv:hep-ph/0606144} \BibitemShut {NoStop}%
\bibitem [{\citenamefont {Babu}\ \emph {et~al.}(2013)\citenamefont {Babu}, \citenamefont {Bhupal~Dev}, \citenamefont {Fortes},\ and\ \citenamefont {Mohapatra}}]{Babu:2013yca}%
  \BibitemOpen
  \bibfield  {author} {\bibinfo {author} {\bibfnamefont {K.~S.}\ \bibnamefont {Babu}}, \bibinfo {author} {\bibfnamefont {P.~S.}\ \bibnamefont {Bhupal~Dev}}, \bibinfo {author} {\bibfnamefont {E.~C. F.~S.}\ \bibnamefont {Fortes}}, \ and\ \bibinfo {author} {\bibfnamefont {R.~N.}\ \bibnamefont {Mohapatra}},\ }\href {\doibase 10.1103/PhysRevD.87.115019} {\bibfield  {journal} {\bibinfo  {journal} {Phys. Rev. D}\ }\textbf {\bibinfo {volume} {87}},\ \bibinfo {pages} {115019} (\bibinfo {year} {2013})},\ \Eprint {http://arxiv.org/abs/1303.6918} {arXiv:1303.6918 [hep-ph]} \BibitemShut {NoStop}%
\bibitem [{\citenamefont {Hewett}\ and\ \citenamefont {Rizzo}(1989)}]{Hewett:1988xc}%
  \BibitemOpen
  \bibfield  {author} {\bibinfo {author} {\bibfnamefont {J.~L.}\ \bibnamefont {Hewett}}\ and\ \bibinfo {author} {\bibfnamefont {T.~G.}\ \bibnamefont {Rizzo}},\ }\href {\doibase 10.1016/0370-1573(89)90071-9} {\bibfield  {journal} {\bibinfo  {journal} {Phys. Rept.}\ }\textbf {\bibinfo {volume} {183}},\ \bibinfo {pages} {193} (\bibinfo {year} {1989})}\BibitemShut {NoStop}%
\bibitem [{\citenamefont {Mohapatra}\ \emph {et~al.}(2008)\citenamefont {Mohapatra}, \citenamefont {Okada},\ and\ \citenamefont {Yu}}]{Mohapatra:2007af}%
  \BibitemOpen
  \bibfield  {author} {\bibinfo {author} {\bibfnamefont {R.~N.}\ \bibnamefont {Mohapatra}}, \bibinfo {author} {\bibfnamefont {N.}~\bibnamefont {Okada}}, \ and\ \bibinfo {author} {\bibfnamefont {H.-B.}\ \bibnamefont {Yu}},\ }\href {\doibase 10.1103/PhysRevD.77.011701} {\bibfield  {journal} {\bibinfo  {journal} {Phys. Rev. D}\ }\textbf {\bibinfo {volume} {77}},\ \bibinfo {pages} {011701} (\bibinfo {year} {2008})},\ \Eprint {http://arxiv.org/abs/0709.1486} {arXiv:0709.1486 [hep-ph]} \BibitemShut {NoStop}%
\bibitem [{\citenamefont {Buchmuller}\ and\ \citenamefont {Wyler}(1986)}]{Buchmuller:1986iq}%
  \BibitemOpen
  \bibfield  {author} {\bibinfo {author} {\bibfnamefont {W.}~\bibnamefont {Buchmuller}}\ and\ \bibinfo {author} {\bibfnamefont {D.}~\bibnamefont {Wyler}},\ }\href {\doibase 10.1016/0370-2693(86)90771-9} {\bibfield  {journal} {\bibinfo  {journal} {Phys. Lett. B}\ }\textbf {\bibinfo {volume} {177}},\ \bibinfo {pages} {377} (\bibinfo {year} {1986})}\BibitemShut {NoStop}%
\bibitem [{\citenamefont {Senjanovic}\ and\ \citenamefont {Sokorac}(1983)}]{Senjanovic:1982ex}%
  \BibitemOpen
  \bibfield  {author} {\bibinfo {author} {\bibfnamefont {G.}~\bibnamefont {Senjanovic}}\ and\ \bibinfo {author} {\bibfnamefont {A.}~\bibnamefont {Sokorac}},\ }\href {\doibase 10.1007/BF01574858} {\bibfield  {journal} {\bibinfo  {journal} {Z. Phys. C}\ }\textbf {\bibinfo {volume} {20}},\ \bibinfo {pages} {255} (\bibinfo {year} {1983})}\BibitemShut {NoStop}%
\bibitem [{\citenamefont {Zee}(1980)}]{Zee:1980ai}%
  \BibitemOpen
  \bibfield  {author} {\bibinfo {author} {\bibfnamefont {A.}~\bibnamefont {Zee}},\ }\href {\doibase 10.1016/0370-2693(80)90349-4} {\bibfield  {journal} {\bibinfo  {journal} {Phys. Lett. B}\ }\textbf {\bibinfo {volume} {93}},\ \bibinfo {pages} {389} (\bibinfo {year} {1980})},\ \bibinfo {note} {[Erratum: Phys.Lett.B 95, 461 (1980)]}\BibitemShut {NoStop}%
\bibitem [{\citenamefont {Babu}(1988)}]{Babu:1988ki}%
  \BibitemOpen
  \bibfield  {author} {\bibinfo {author} {\bibfnamefont {K.~S.}\ \bibnamefont {Babu}},\ }\href {\doibase 10.1016/0370-2693(88)91584-5} {\bibfield  {journal} {\bibinfo  {journal} {Phys. Lett. B}\ }\textbf {\bibinfo {volume} {203}},\ \bibinfo {pages} {132} (\bibinfo {year} {1988})}\BibitemShut {NoStop}%
\bibitem [{\citenamefont {Pascual-Dias}\ \emph {et~al.}(2020)\citenamefont {Pascual-Dias}, \citenamefont {Saha},\ and\ \citenamefont {London}}]{Pascual-Dias:2020hxo}%
  \BibitemOpen
  \bibfield  {author} {\bibinfo {author} {\bibfnamefont {B.}~\bibnamefont {Pascual-Dias}}, \bibinfo {author} {\bibfnamefont {P.}~\bibnamefont {Saha}}, \ and\ \bibinfo {author} {\bibfnamefont {D.}~\bibnamefont {London}},\ }\href {\doibase 10.1007/JHEP07(2020)144} {\bibfield  {journal} {\bibinfo  {journal} {JHEP}\ }\textbf {\bibinfo {volume} {07}},\ \bibinfo {pages} {144} (\bibinfo {year} {2020})},\ \Eprint {http://arxiv.org/abs/2006.13385} {arXiv:2006.13385 [hep-ph]} \BibitemShut {NoStop}%
\bibitem [{\citenamefont {Giudice}\ \emph {et~al.}(2011)\citenamefont {Giudice}, \citenamefont {Gripaios},\ and\ \citenamefont {Sundrum}}]{Giudice:2011ak}%
  \BibitemOpen
  \bibfield  {author} {\bibinfo {author} {\bibfnamefont {G.~F.}\ \bibnamefont {Giudice}}, \bibinfo {author} {\bibfnamefont {B.}~\bibnamefont {Gripaios}}, \ and\ \bibinfo {author} {\bibfnamefont {R.}~\bibnamefont {Sundrum}},\ }\href {\doibase 10.1007/JHEP08(2011)055} {\bibfield  {journal} {\bibinfo  {journal} {JHEP}\ }\textbf {\bibinfo {volume} {08}},\ \bibinfo {pages} {055} (\bibinfo {year} {2011})},\ \Eprint {http://arxiv.org/abs/1105.3161} {arXiv:1105.3161 [hep-ph]} \BibitemShut {NoStop}%
\bibitem [{\citenamefont {Dor{\v{s}}ner}\ \emph {et~al.}(2016)\citenamefont {Dor{\v{s}}ner}, \citenamefont {Fajfer}, \citenamefont {Greljo}, \citenamefont {Kamenik},\ and\ \citenamefont {Ko{\v{s}}nik}}]{Dorsner:2016wpm}%
  \BibitemOpen
  \bibfield  {author} {\bibinfo {author} {\bibfnamefont {I.}~\bibnamefont {Dor{\v{s}}ner}}, \bibinfo {author} {\bibfnamefont {S.}~\bibnamefont {Fajfer}}, \bibinfo {author} {\bibfnamefont {A.}~\bibnamefont {Greljo}}, \bibinfo {author} {\bibfnamefont {J.~F.}\ \bibnamefont {Kamenik}}, \ and\ \bibinfo {author} {\bibfnamefont {N.}~\bibnamefont {Ko{\v{s}}nik}},\ }\href {\doibase 10.1016/j.physrep.2016.06.001} {\bibfield  {journal} {\bibinfo  {journal} {Phys. Rept.}\ }\textbf {\bibinfo {volume} {641}},\ \bibinfo {pages} {1} (\bibinfo {year} {2016})},\ \Eprint {http://arxiv.org/abs/1603.04993} {arXiv:1603.04993 [hep-ph]} \BibitemShut {NoStop}%
\bibitem [{\citenamefont {Crivellin}\ and\ \citenamefont {Schnell}(2022)}]{Crivellin:2021ejk}%
  \BibitemOpen
  \bibfield  {author} {\bibinfo {author} {\bibfnamefont {A.}~\bibnamefont {Crivellin}}\ and\ \bibinfo {author} {\bibfnamefont {L.}~\bibnamefont {Schnell}},\ }\href {\doibase 10.1016/j.cpc.2021.108188} {\bibfield  {journal} {\bibinfo  {journal} {Comput. Phys. Commun.}\ }\textbf {\bibinfo {volume} {271}},\ \bibinfo {pages} {108188} (\bibinfo {year} {2022})},\ \Eprint {http://arxiv.org/abs/2105.04844} {arXiv:2105.04844 [hep-ph]} \BibitemShut {NoStop}%
\bibitem [{\citenamefont {Lavelle}\ and\ \citenamefont {McMullan}(1997)}]{Lavelle:1995ty}%
  \BibitemOpen
  \bibfield  {author} {\bibinfo {author} {\bibfnamefont {M.}~\bibnamefont {Lavelle}}\ and\ \bibinfo {author} {\bibfnamefont {D.}~\bibnamefont {McMullan}},\ }\href {\doibase 10.1016/S0370-1573(96)00019-1} {\bibfield  {journal} {\bibinfo  {journal} {Phys. Rept.}\ }\textbf {\bibinfo {volume} {279}},\ \bibinfo {pages} {1} (\bibinfo {year} {1997})},\ \Eprint {http://arxiv.org/abs/hep-ph/9509344} {arXiv:hep-ph/9509344} \BibitemShut {NoStop}%
\bibitem [{\citenamefont {Navas}\ \emph {et~al.}(2024{\natexlab{a}})\citenamefont {Navas} \emph {et~al.}}]{ParticleDataGroupConstituentQuarks:2024cfk}%
  \BibitemOpen
  \bibfield  {author} {\bibinfo {author} {\bibfnamefont {S.}~\bibnamefont {Navas}} \emph {et~al.} (\bibinfo {collaboration} {Particle Data Group, Volume I Chapter 60: Quark Masses}),\ }\href {\doibase 10.1103/PhysRevD.110.030001} {\bibfield  {journal} {\bibinfo  {journal} {Phys. Rev. D}\ }\textbf {\bibinfo {volume} {110}},\ \bibinfo {pages} {030001} (\bibinfo {year} {2024}{\natexlab{a}})}\BibitemShut {NoStop}%
\bibitem [{\citenamefont {Bigaran}\ and\ \citenamefont {Volkas}(2020)}]{Bigaran:2020jil}%
  \BibitemOpen
  \bibfield  {author} {\bibinfo {author} {\bibfnamefont {I.}~\bibnamefont {Bigaran}}\ and\ \bibinfo {author} {\bibfnamefont {R.~R.}\ \bibnamefont {Volkas}},\ }\href {\doibase 10.1103/PhysRevD.102.075037} {\bibfield  {journal} {\bibinfo  {journal} {Phys. Rev. D}\ }\textbf {\bibinfo {volume} {102}},\ \bibinfo {pages} {075037} (\bibinfo {year} {2020})},\ \Eprint {http://arxiv.org/abs/2002.12544} {arXiv:2002.12544 [hep-ph]} \BibitemShut {NoStop}%
\bibitem [{\citenamefont {McLaughlin}\ and\ \citenamefont {Ng}(1999)}]{McLaughlin:1999rr}%
  \BibitemOpen
  \bibfield  {author} {\bibinfo {author} {\bibfnamefont {G.~C.}\ \bibnamefont {McLaughlin}}\ and\ \bibinfo {author} {\bibfnamefont {J.~N.}\ \bibnamefont {Ng}},\ }\href {\doibase 10.1016/S0370-2693(99)00490-6} {\bibfield  {journal} {\bibinfo  {journal} {Phys. Lett. B}\ }\textbf {\bibinfo {volume} {455}},\ \bibinfo {pages} {224} (\bibinfo {year} {1999})},\ \Eprint {http://arxiv.org/abs/hep-ph/9903509} {arXiv:hep-ph/9903509} \BibitemShut {NoStop}%
\bibitem [{\citenamefont {Baldini}\ \emph {et~al.}(2016)\citenamefont {Baldini} \emph {et~al.}}]{MEG:2016leq}%
  \BibitemOpen
  \bibfield  {author} {\bibinfo {author} {\bibfnamefont {A.~M.}\ \bibnamefont {Baldini}} \emph {et~al.} (\bibinfo {collaboration} {MEG}),\ }\href {\doibase 10.1140/epjc/s10052-016-4271-x} {\bibfield  {journal} {\bibinfo  {journal} {Eur. Phys. J. C}\ }\textbf {\bibinfo {volume} {76}},\ \bibinfo {pages} {434} (\bibinfo {year} {2016})},\ \Eprint {http://arxiv.org/abs/1605.05081} {arXiv:1605.05081 [hep-ex]} \BibitemShut {NoStop}%
\bibitem [{\citenamefont {Afanaciev}\ \emph {et~al.}(2024)\citenamefont {Afanaciev} \emph {et~al.}}]{MEGII:2023ltw}%
  \BibitemOpen
  \bibfield  {author} {\bibinfo {author} {\bibfnamefont {K.}~\bibnamefont {Afanaciev}} \emph {et~al.} (\bibinfo {collaboration} {MEG II}),\ }\href {\doibase 10.1140/epjc/s10052-024-12416-2} {\bibfield  {journal} {\bibinfo  {journal} {Eur. Phys. J. C}\ }\textbf {\bibinfo {volume} {84}},\ \bibinfo {pages} {216} (\bibinfo {year} {2024})},\ \bibinfo {note} {[Erratum: Eur.Phys.J.C 84, 1042 (2024)]},\ \Eprint {http://arxiv.org/abs/2310.12614} {arXiv:2310.12614 [hep-ex]} \BibitemShut {NoStop}%
\bibitem [{\citenamefont {Aad}\ \emph {et~al.}(2021{\natexlab{a}})\citenamefont {Aad} \emph {et~al.}}]{ATLAS:2021oiz}%
  \BibitemOpen
  \bibfield  {author} {\bibinfo {author} {\bibfnamefont {G.}~\bibnamefont {Aad}} \emph {et~al.} (\bibinfo {collaboration} {ATLAS}),\ }\href {\doibase 10.1007/JHEP06(2021)179} {\bibfield  {journal} {\bibinfo  {journal} {JHEP}\ }\textbf {\bibinfo {volume} {06}},\ \bibinfo {pages} {179} (\bibinfo {year} {2021}{\natexlab{a}})},\ \Eprint {http://arxiv.org/abs/2101.11582} {arXiv:2101.11582 [hep-ex]} \BibitemShut {NoStop}%
\bibitem [{\citenamefont {Tumasyan}\ \emph {et~al.}(2022)\citenamefont {Tumasyan} \emph {et~al.}}]{CMS:2022nty}%
  \BibitemOpen
  \bibfield  {author} {\bibinfo {author} {\bibfnamefont {A.}~\bibnamefont {Tumasyan}} \emph {et~al.} (\bibinfo {collaboration} {CMS}),\ }\href {\doibase 10.1103/PhysRevD.105.112007} {\bibfield  {journal} {\bibinfo  {journal} {Phys. Rev. D}\ }\textbf {\bibinfo {volume} {105}},\ \bibinfo {pages} {112007} (\bibinfo {year} {2022})},\ \Eprint {http://arxiv.org/abs/2202.08676} {arXiv:2202.08676 [hep-ex]} \BibitemShut {NoStop}%
\bibitem [{\citenamefont {Aad}\ \emph {et~al.}(2021{\natexlab{b}})\citenamefont {Aad} \emph {et~al.}}]{ATLAS:2020xov}%
  \BibitemOpen
  \bibfield  {author} {\bibinfo {author} {\bibfnamefont {G.}~\bibnamefont {Aad}} \emph {et~al.} (\bibinfo {collaboration} {ATLAS}),\ }\href {\doibase 10.1140/epjc/s10052-021-09009-8} {\bibfield  {journal} {\bibinfo  {journal} {Eur. Phys. J. C}\ }\textbf {\bibinfo {volume} {81}},\ \bibinfo {pages} {313} (\bibinfo {year} {2021}{\natexlab{b}})},\ \Eprint {http://arxiv.org/abs/2010.02098} {arXiv:2010.02098 [hep-ex]} \BibitemShut {NoStop}%
\bibitem [{\citenamefont {Navas}\ \emph {et~al.}(2024{\natexlab{b}})\citenamefont {Navas} \emph {et~al.}}]{ParticleDataGroupLeptoquarks:2024cfk}%
  \BibitemOpen
  \bibfield  {author} {\bibinfo {author} {\bibfnamefont {S.}~\bibnamefont {Navas}} \emph {et~al.} (\bibinfo {collaboration} {Particle Data Group, Volume I Chapter 94: Leptoquarks}),\ }\href {\doibase 10.1103/PhysRevD.110.030001} {\bibfield  {journal} {\bibinfo  {journal} {Phys. Rev. D}\ }\textbf {\bibinfo {volume} {110}},\ \bibinfo {pages} {030001} (\bibinfo {year} {2024}{\natexlab{b}})}\BibitemShut {NoStop}%
\bibitem [{\citenamefont {Herrero-Garcia}\ \emph {et~al.}(2014)\citenamefont {Herrero-Garcia}, \citenamefont {Nebot}, \citenamefont {Rius},\ and\ \citenamefont {Santamaria}}]{Herrero-Garcia:2014hfa}%
  \BibitemOpen
  \bibfield  {author} {\bibinfo {author} {\bibfnamefont {J.}~\bibnamefont {Herrero-Garcia}}, \bibinfo {author} {\bibfnamefont {M.}~\bibnamefont {Nebot}}, \bibinfo {author} {\bibfnamefont {N.}~\bibnamefont {Rius}}, \ and\ \bibinfo {author} {\bibfnamefont {A.}~\bibnamefont {Santamaria}},\ }\href {\doibase 10.1016/j.nuclphysb.2014.06.001} {\bibfield  {journal} {\bibinfo  {journal} {Nucl. Phys. B}\ }\textbf {\bibinfo {volume} {885}},\ \bibinfo {pages} {542} (\bibinfo {year} {2014})},\ \Eprint {http://arxiv.org/abs/1402.4491} {arXiv:1402.4491 [hep-ph]} \BibitemShut {NoStop}%
\bibitem [{\citenamefont {Pich}(2014)}]{Pich:2013lsa}%
  \BibitemOpen
  \bibfield  {author} {\bibinfo {author} {\bibfnamefont {A.}~\bibnamefont {Pich}},\ }\href {\doibase 10.1016/j.ppnp.2013.11.002} {\bibfield  {journal} {\bibinfo  {journal} {Prog. Part. Nucl. Phys.}\ }\textbf {\bibinfo {volume} {75}},\ \bibinfo {pages} {41} (\bibinfo {year} {2014})},\ \Eprint {http://arxiv.org/abs/1310.7922} {arXiv:1310.7922 [hep-ph]} \BibitemShut {NoStop}%
\bibitem [{ATL(2019)}]{ATLAS-CONF-2019-007}%
  \BibitemOpen
  \href {https://cds.cern.ch/record/2668385} {\emph {\bibinfo {title} {{Search for New Phenomena in Dijet Events using 139 fb$^{−1}$ of $pp$ collisions at $\sqrt{s}$ = 13TeV collected with the ATLAS Detector}}}},\ \bibinfo {type} {Tech. Rep.}\ (\bibinfo  {institution} {CERN},\ \bibinfo {address} {Geneva},\ \bibinfo {year} {2019})\BibitemShut {NoStop}%
\bibitem [{\citenamefont {Sirunyan}\ \emph {et~al.}(2020)\citenamefont {Sirunyan} \emph {et~al.}}]{CMS:2019gwf}%
  \BibitemOpen
  \bibfield  {author} {\bibinfo {author} {\bibfnamefont {A.~M.}\ \bibnamefont {Sirunyan}} \emph {et~al.} (\bibinfo {collaboration} {CMS}),\ }\href {\doibase 10.1007/JHEP05(2020)033} {\bibfield  {journal} {\bibinfo  {journal} {JHEP}\ }\textbf {\bibinfo {volume} {05}},\ \bibinfo {pages} {033} (\bibinfo {year} {2020})},\ \Eprint {http://arxiv.org/abs/1911.03947} {arXiv:1911.03947 [hep-ex]} \BibitemShut {NoStop}%
\bibitem [{\citenamefont {Dong}\ \emph {et~al.}(2012)\citenamefont {Dong}, \citenamefont {Durieux}, \citenamefont {Gerard}, \citenamefont {Han},\ and\ \citenamefont {Maltoni}}]{Dong:2011rh}%
  \BibitemOpen
  \bibfield  {author} {\bibinfo {author} {\bibfnamefont {Z.}~\bibnamefont {Dong}}, \bibinfo {author} {\bibfnamefont {G.}~\bibnamefont {Durieux}}, \bibinfo {author} {\bibfnamefont {J.-M.}\ \bibnamefont {Gerard}}, \bibinfo {author} {\bibfnamefont {T.}~\bibnamefont {Han}}, \ and\ \bibinfo {author} {\bibfnamefont {F.}~\bibnamefont {Maltoni}},\ }\href {\doibase 10.1103/PhysRevD.85.016006} {\bibfield  {journal} {\bibinfo  {journal} {Phys. Rev. D}\ }\textbf {\bibinfo {volume} {85}},\ \bibinfo {pages} {016006} (\bibinfo {year} {2012})},\ \Eprint {http://arxiv.org/abs/1107.3805} {arXiv:1107.3805 [hep-ph]} \BibitemShut {NoStop}%
\bibitem [{\citenamefont {Hayrapetyan}\ \emph {et~al.}(2024)\citenamefont {Hayrapetyan} \emph {et~al.}}]{CMS:2024dzv}%
  \BibitemOpen
  \bibfield  {author} {\bibinfo {author} {\bibfnamefont {A.}~\bibnamefont {Hayrapetyan}} \emph {et~al.} (\bibinfo {collaboration} {CMS}),\ }\href {\doibase 10.1103/PhysRevLett.132.241802} {\bibfield  {journal} {\bibinfo  {journal} {Phys. Rev. Lett.}\ }\textbf {\bibinfo {volume} {132}},\ \bibinfo {pages} {241802} (\bibinfo {year} {2024})},\ \Eprint {http://arxiv.org/abs/2402.18461} {arXiv:2402.18461 [hep-ex]} \BibitemShut {NoStop}%
\bibitem [{\citenamefont {Rehermann}\ and\ \citenamefont {Tweedie}(2011)}]{Rehermann:2010vq}%
  \BibitemOpen
  \bibfield  {author} {\bibinfo {author} {\bibfnamefont {K.}~\bibnamefont {Rehermann}}\ and\ \bibinfo {author} {\bibfnamefont {B.}~\bibnamefont {Tweedie}},\ }\href {\doibase 10.1007/JHEP03(2011)059} {\bibfield  {journal} {\bibinfo  {journal} {JHEP}\ }\textbf {\bibinfo {volume} {03}},\ \bibinfo {pages} {059} (\bibinfo {year} {2011})},\ \Eprint {http://arxiv.org/abs/1007.2221} {arXiv:1007.2221 [hep-ph]} \BibitemShut {NoStop}%
\bibitem [{\citenamefont {Aad}\ \emph {et~al.}(2011)\citenamefont {Aad} \emph {et~al.}}]{ATLAS:2011izm}%
  \BibitemOpen
  \bibfield  {author} {\bibinfo {author} {\bibfnamefont {G.}~\bibnamefont {Aad}} \emph {et~al.} (\bibinfo {collaboration} {ATLAS}),\ }\href {\doibase 10.1007/JHEP10(2011)107} {\bibfield  {journal} {\bibinfo  {journal} {JHEP}\ }\textbf {\bibinfo {volume} {10}},\ \bibinfo {pages} {107} (\bibinfo {year} {2011})},\ \Eprint {http://arxiv.org/abs/1108.0366} {arXiv:1108.0366 [hep-ex]} \BibitemShut {NoStop}%
\bibitem [{\citenamefont {Khachatryan}\ \emph {et~al.}(2016)\citenamefont {Khachatryan} \emph {et~al.}}]{CMS:2016mku}%
  \BibitemOpen
  \bibfield  {author} {\bibinfo {author} {\bibfnamefont {V.}~\bibnamefont {Khachatryan}} \emph {et~al.} (\bibinfo {collaboration} {CMS}),\ }\href {\doibase 10.1140/epjc/s10052-016-4261-z} {\bibfield  {journal} {\bibinfo  {journal} {Eur. Phys. J. C}\ }\textbf {\bibinfo {volume} {76}},\ \bibinfo {pages} {439} (\bibinfo {year} {2016})},\ \Eprint {http://arxiv.org/abs/1605.03171} {arXiv:1605.03171 [hep-ex]} \BibitemShut {NoStop}%
\bibitem [{\citenamefont {Alwall}\ \emph {et~al.}(2014)\citenamefont {Alwall}, \citenamefont {Frederix}, \citenamefont {Frixione}, \citenamefont {Hirschi}, \citenamefont {Maltoni}, \citenamefont {Mattelaer}, \citenamefont {Shao}, \citenamefont {Stelzer}, \citenamefont {Torrielli},\ and\ \citenamefont {Zaro}}]{Alwall:2014hca}%
  \BibitemOpen
  \bibfield  {author} {\bibinfo {author} {\bibfnamefont {J.}~\bibnamefont {Alwall}}, \bibinfo {author} {\bibfnamefont {R.}~\bibnamefont {Frederix}}, \bibinfo {author} {\bibfnamefont {S.}~\bibnamefont {Frixione}}, \bibinfo {author} {\bibfnamefont {V.}~\bibnamefont {Hirschi}}, \bibinfo {author} {\bibfnamefont {F.}~\bibnamefont {Maltoni}}, \bibinfo {author} {\bibfnamefont {O.}~\bibnamefont {Mattelaer}}, \bibinfo {author} {\bibfnamefont {H.~S.}\ \bibnamefont {Shao}}, \bibinfo {author} {\bibfnamefont {T.}~\bibnamefont {Stelzer}}, \bibinfo {author} {\bibfnamefont {P.}~\bibnamefont {Torrielli}}, \ and\ \bibinfo {author} {\bibfnamefont {M.}~\bibnamefont {Zaro}},\ }\href {\doibase 10.1007/JHEP07(2014)079} {\bibfield  {journal} {\bibinfo  {journal} {JHEP}\ }\textbf {\bibinfo {volume} {07}},\ \bibinfo {pages} {079} (\bibinfo {year} {2014})},\ \Eprint {http://arxiv.org/abs/1405.0301} {arXiv:1405.0301 [hep-ph]} \BibitemShut {NoStop}%
\bibitem [{\citenamefont {Luty}(1992)}]{Luty:1992un}%
  \BibitemOpen
  \bibfield  {author} {\bibinfo {author} {\bibfnamefont {M.~A.}\ \bibnamefont {Luty}},\ }\href {\doibase 10.1103/PhysRevD.45.455} {\bibfield  {journal} {\bibinfo  {journal} {Phys. Rev. D}\ }\textbf {\bibinfo {volume} {45}},\ \bibinfo {pages} {455} (\bibinfo {year} {1992})}\BibitemShut {NoStop}%
\bibitem [{\citenamefont {Gonzalez-Garcia}\ and\ \citenamefont {Yokoyama}()}]{pdg_pmns_matrix}%
  \BibitemOpen
  \bibfield  {author} {\bibinfo {author} {\bibfnamefont {M.}~\bibnamefont {Gonzalez-Garcia}}\ and\ \bibinfo {author} {\bibfnamefont {M.}~\bibnamefont {Yokoyama}},\ }\href@noop {} {}\bibinfo {howpublished} {\url{https://pdg.lbl.gov/2024/reviews/rpp2024-rev-neutrino-mixing.pdf}},\ \bibinfo {note} {[Accessed 25-08-2025]}\BibitemShut {NoStop}%
\bibitem [{\citenamefont {Pontecorvo}(1958)}]{Pontecorvo:1957qd}%
  \BibitemOpen
  \bibfield  {author} {\bibinfo {author} {\bibfnamefont {B.}~\bibnamefont {Pontecorvo}},\ }\href@noop {} {\bibfield  {journal} {\bibinfo  {journal} {Sov. Phys. JETP}\ }\textbf {\bibinfo {volume} {7}},\ \bibinfo {pages} {172} (\bibinfo {year} {1958})}\BibitemShut {NoStop}%
\bibitem [{\citenamefont {Maki}\ \emph {et~al.}(1962)\citenamefont {Maki}, \citenamefont {Nakagawa},\ and\ \citenamefont {Sakata}}]{Maki:1962mu}%
  \BibitemOpen
  \bibfield  {author} {\bibinfo {author} {\bibfnamefont {Z.}~\bibnamefont {Maki}}, \bibinfo {author} {\bibfnamefont {M.}~\bibnamefont {Nakagawa}}, \ and\ \bibinfo {author} {\bibfnamefont {S.}~\bibnamefont {Sakata}},\ }\href {\doibase 10.1143/PTP.28.870} {\bibfield  {journal} {\bibinfo  {journal} {Prog. Theor. Phys.}\ }\textbf {\bibinfo {volume} {28}},\ \bibinfo {pages} {870} (\bibinfo {year} {1962})}\BibitemShut {NoStop}%
\bibitem [{\citenamefont {Ceccucci}\ \emph {et~al.}()\citenamefont {Ceccucci}, \citenamefont {Ligeti},\ and\ \citenamefont {Sakai}}]{pdg_ckm_matrix}%
  \BibitemOpen
  \bibfield  {author} {\bibinfo {author} {\bibfnamefont {A.}~\bibnamefont {Ceccucci}}, \bibinfo {author} {\bibfnamefont {Z.}~\bibnamefont {Ligeti}}, \ and\ \bibinfo {author} {\bibfnamefont {Y.}~\bibnamefont {Sakai}},\ }\href@noop {} {}\bibinfo {howpublished} {\url{https://pdg.lbl.gov/2019/reviews/rpp2019-rev-ckm-matrix.pdf}},\ \bibinfo {note} {[Accessed 20-08-2025]}\BibitemShut {NoStop}%
\bibitem [{\citenamefont {Cabibbo}(1963)}]{Cabibbo:1963yz}%
  \BibitemOpen
  \bibfield  {author} {\bibinfo {author} {\bibfnamefont {N.}~\bibnamefont {Cabibbo}},\ }\href {\doibase 10.1103/PhysRevLett.10.531} {\bibfield  {journal} {\bibinfo  {journal} {Phys. Rev. Lett.}\ }\textbf {\bibinfo {volume} {10}},\ \bibinfo {pages} {531} (\bibinfo {year} {1963})}\BibitemShut {NoStop}%
\bibitem [{\citenamefont {Kobayashi}\ and\ \citenamefont {Maskawa}(1973)}]{Kobayashi:1973fv}%
  \BibitemOpen
  \bibfield  {author} {\bibinfo {author} {\bibfnamefont {M.}~\bibnamefont {Kobayashi}}\ and\ \bibinfo {author} {\bibfnamefont {T.}~\bibnamefont {Maskawa}},\ }\href {\doibase 10.1143/PTP.49.652} {\bibfield  {journal} {\bibinfo  {journal} {Prog. Theor. Phys.}\ }\textbf {\bibinfo {volume} {49}},\ \bibinfo {pages} {652} (\bibinfo {year} {1973})}\BibitemShut {NoStop}%
\bibitem [{\citenamefont {Benedikt}\ \emph {et~al.}(2025)\citenamefont {Benedikt} \emph {et~al.}}]{FCC:2025lpp}%
  \BibitemOpen
  \bibfield  {author} {\bibinfo {author} {\bibfnamefont {M.}~\bibnamefont {Benedikt}} \emph {et~al.} (\bibinfo {collaboration} {FCC}),\ }\href {\doibase 10.17181/CERN.9DKX.TDH9} {\  (\bibinfo {year} {2025}),\ 10.17181/CERN.9DKX.TDH9},\ \Eprint {http://arxiv.org/abs/2505.00272} {arXiv:2505.00272 [hep-ex]} \BibitemShut {NoStop}%
\bibitem [{\citenamefont {Dong}\ \emph {et~al.}(2018)\citenamefont {Dong} \emph {et~al.}}]{CEPCStudyGroup:2018ghi}%
  \BibitemOpen
  \bibfield  {author} {\bibinfo {author} {\bibfnamefont {M.}~\bibnamefont {Dong}} \emph {et~al.} (\bibinfo {collaboration} {CEPC Study Group}),\ }\href@noop {} {\  (\bibinfo {year} {2018})},\ \Eprint {http://arxiv.org/abs/1811.10545} {arXiv:1811.10545 [hep-ex]} \BibitemShut {NoStop}%
\bibitem [{\citenamefont {Ahmad}\ \emph {et~al.}(2015)\citenamefont {Ahmad} \emph {et~al.}}]{CEPC-SPPCStudyGroup:2015csa}%
  \BibitemOpen
  \bibfield  {author} {\bibinfo {author} {\bibfnamefont {M.}~\bibnamefont {Ahmad}} \emph {et~al.} (\bibinfo {collaboration} {CEPC-SPPC Study Group}),\ }\href@noop {} {\  (\bibinfo {year} {2015})}\BibitemShut {NoStop}%
\bibitem [{\citenamefont {de~Blas}\ \emph {et~al.}(2018)\citenamefont {de~Blas} \emph {et~al.}}]{CLIC:2018fvx}%
  \BibitemOpen
  \bibfield  {author} {\bibinfo {author} {\bibfnamefont {J.}~\bibnamefont {de~Blas}} \emph {et~al.} (\bibinfo {collaboration} {CLIC}),\ }\href {\doibase 10.23731/CYRM-2018-003} {\bibfield  {journal} {\bibinfo  {journal} {CERN Yellow Rep. Monogr.}\ }\textbf {\bibinfo {volume} {3}},\ \bibinfo {pages} {1} (\bibinfo {year} {2018})},\ \Eprint {http://arxiv.org/abs/1812.02093} {arXiv:1812.02093 [hep-ph]} \BibitemShut {NoStop}%
\bibitem [{ILC(2013)}]{ILC:2013jhg}%
  \BibitemOpen
  \href@noop {} {\  (\bibinfo {year} {2013})},\ \Eprint {http://arxiv.org/abs/1306.6352} {arXiv:1306.6352 [hep-ph]} \BibitemShut {NoStop}%
\bibitem [{\citenamefont {Accettura}\ \emph {et~al.}(2023)\citenamefont {Accettura} \emph {et~al.}}]{Accettura:2023ked}%
  \BibitemOpen
  \bibfield  {author} {\bibinfo {author} {\bibfnamefont {C.}~\bibnamefont {Accettura}} \emph {et~al.},\ }\href {\doibase 10.1140/epjc/s10052-023-11889-x} {\bibfield  {journal} {\bibinfo  {journal} {Eur. Phys. J. C}\ }\textbf {\bibinfo {volume} {83}},\ \bibinfo {pages} {864} (\bibinfo {year} {2023})},\ \bibinfo {note} {[Erratum: Eur.Phys.J.C 84, 36 (2024)]},\ \Eprint {http://arxiv.org/abs/2303.08533} {arXiv:2303.08533 [physics.acc-ph]} \BibitemShut {NoStop}%
\bibitem [{\citenamefont {Accettura}\ \emph {et~al.}(2025)\citenamefont {Accettura} \emph {et~al.}}]{InternationalMuonCollider:2025sys}%
  \BibitemOpen
  \bibfield  {author} {\bibinfo {author} {\bibfnamefont {C.}~\bibnamefont {Accettura}} \emph {et~al.} (\bibinfo {collaboration} {International Muon Collider}),\ }\href@noop {} {\  (\bibinfo {year} {2025})},\ \Eprint {http://arxiv.org/abs/2504.21417} {arXiv:2504.21417 [physics.acc-ph]} \BibitemShut {NoStop}%
\bibitem [{\citenamefont {Goodsell}\ and\ \citenamefont {Staub}(2018)}]{Goodsell:2018tti}%
  \BibitemOpen
  \bibfield  {author} {\bibinfo {author} {\bibfnamefont {M.~D.}\ \bibnamefont {Goodsell}}\ and\ \bibinfo {author} {\bibfnamefont {F.}~\bibnamefont {Staub}},\ }\href {\doibase 10.1140/epjc/s10052-018-6127-z} {\bibfield  {journal} {\bibinfo  {journal} {Eur. Phys. J. C}\ }\textbf {\bibinfo {volume} {78}},\ \bibinfo {pages} {649} (\bibinfo {year} {2018})},\ \Eprint {http://arxiv.org/abs/1805.07306} {arXiv:1805.07306 [hep-ph]} \BibitemShut {NoStop}%
\bibitem [{\citenamefont {Aoki}\ \emph {et~al.}(2017)\citenamefont {Aoki}, \citenamefont {Izubuchi}, \citenamefont {Shintani},\ and\ \citenamefont {Soni}}]{Aoki:2017puj}%
  \BibitemOpen
  \bibfield  {author} {\bibinfo {author} {\bibfnamefont {Y.}~\bibnamefont {Aoki}}, \bibinfo {author} {\bibfnamefont {T.}~\bibnamefont {Izubuchi}}, \bibinfo {author} {\bibfnamefont {E.}~\bibnamefont {Shintani}}, \ and\ \bibinfo {author} {\bibfnamefont {A.}~\bibnamefont {Soni}},\ }\href {\doibase 10.1103/PhysRevD.96.014506} {\bibfield  {journal} {\bibinfo  {journal} {Phys. Rev. D}\ }\textbf {\bibinfo {volume} {96}},\ \bibinfo {pages} {014506} (\bibinfo {year} {2017})},\ \Eprint {http://arxiv.org/abs/1705.01338} {arXiv:1705.01338 [hep-lat]} \BibitemShut {NoStop}%
\bibitem [{\citenamefont {Hou}\ \emph {et~al.}(2005)\citenamefont {Hou}, \citenamefont {Nagashima},\ and\ \citenamefont {Soddu}}]{Hou:2005iu}%
  \BibitemOpen
  \bibfield  {author} {\bibinfo {author} {\bibfnamefont {W.-S.}\ \bibnamefont {Hou}}, \bibinfo {author} {\bibfnamefont {M.}~\bibnamefont {Nagashima}}, \ and\ \bibinfo {author} {\bibfnamefont {A.}~\bibnamefont {Soddu}},\ }\href {\doibase 10.1103/PhysRevD.72.095001} {\bibfield  {journal} {\bibinfo  {journal} {Phys. Rev. D}\ }\textbf {\bibinfo {volume} {72}},\ \bibinfo {pages} {095001} (\bibinfo {year} {2005})},\ \Eprint {http://arxiv.org/abs/hep-ph/0509006} {arXiv:hep-ph/0509006} \BibitemShut {NoStop}%
\end{thebibliography}%

\end{document}